\documentclass[useAMS,usenatbib]{mn2e}

 \pdfoutput=1
\usepackage{natbib}
\usepackage{verbatim} 
\usepackage{amsmath}
\usepackage{amssymb}
\usepackage{textcomp}
\usepackage{longtable}
\usepackage{graphicx}
\usepackage{float}
\def\jnl@style{\it}
\def\aaref@jnl#1{{\jnl@style#1}}
\def\aaref@jnl#1{{\jnl@style#1}}

\def\aj{\aaref@jnl{AJ}}                   % Astronomical Journal
\def\araa{\aaref@jnl{ARA\&A}}             % Annual Review of Astron and Astrophys
\def\apj{\aaref@jnl{ApJ}}                 % Astrophysical Journal
\def\apjl{\aaref@jnl{ApJ}}                % Astrophysical Journal, Letters
\def\apjs{\aaref@jnl{ApJS}}               % Astrophysical Journal, Supplement
\def\ao{\aaref@jnl{Appl.~Opt.}}           % Applied Optics
\def\apss{\aaref@jnl{Ap\&SS}}             % Astrophysics and Space Science
\def\aap{\aaref@jnl{A\&A}}                % Astronomy and Astrophysics
\def\aapr{\aaref@jnl{A\&A~Rev.}}          % Astronomy and Astrophysics Reviews
\def\aaps{\aaref@jnl{A\&AS}}              % Astronomy and Astrophysics, Supplement
\def\azh{\aaref@jnl{AZh}}                 % Astronomicheskii Zhurnal
\def\baas{\aaref@jnl{BAAS}}               % Bulletin of the AAS
\def\jrasc{\aaref@jnl{JRASC}}             % Journal of the RAS of Canada
\def\memras{\aaref@jnl{MmRAS}}            % Memoirs of the RAS
\def\mnras{\aaref@jnl{MNRAS}}             % Monthly Notices of the RAS
\def\pra{\aaref@jnl{Phys.~Rev.~A}}        % Physical Review A: General Physics
\def\prb{\aaref@jnl{Phys.~Rev.~B}}        % Physical Review B: Solid State
\def\prc{\aaref@jnl{Phys.~Rev.~C}}        % Physical Review C
\def\prd{\aaref@jnl{Phys.~Rev.~D}}        % Physical Review D
\def\pre{\aaref@jnl{Phys.~Rev.~E}}        % Physical Review E
\def\prl{\aaref@jnl{Phys.~Rev.~Lett.}}    % Physical Review Letters
\def\pasp{\aaref@jnl{PASP}}               % Publications of the ASP
\def\pasj{\aaref@jnl{PASJ}}               % Publications of the ASJ
\def\qjras{\aaref@jnl{QJRAS}}             % Quarterly Journal of the RAS
\def\skytel{\aaref@jnl{S\&T}}             % Sky and Telescope
\def\solphys{\aaref@jnl{Sol.~Phys.}}      % Solar Physics
\def\sovast{\aaref@jnl{Soviet~Ast.}}      % Soviet Astronomy
\def\ssr{\aaref@jnl{Space~Sci.~Rev.}}     % Space Science Reviews
\def\zap{\aaref@jnl{ZAp}}                 % Zeitschrift fuer Astrophysik
\def\nat{\aaref@jnl{Nature}}              % Nature
\def\iaucirc{\aaref@jnl{IAU~Circ.}}       % IAU Cirulars
\def\aplett{\aaref@jnl{Astrophys.~Lett.}} % Astrophysics Letters
\def\apspr{\aaref@jnl{Astrophys.~Space~Phys.~Res.}}
\def\bain{\aaref@jnl{Bull.~Astron.~Inst.~Netherlands}} 
\def\fcp{\aaref@jnl{Fund.~Cosmic~Phys.}}  % Fundamental Cosmic Physics
\def\gca{\aaref@jnl{Geochim.~Cosmochim.~Acta}}   % Geochimica Cosmochimica Acta
\def\grl{\aaref@jnl{Geophys.~Res.~Lett.}} % Geophysics Research Letters
\def\jcp{\aaref@jnl{J.~Chem.~Phys.}}      % Journal of Chemical Physics
\def\jgr{\aaref@jnl{J.~Geophys.~Res.}}    % Journal of Geophysics Research
\def\jqsrt{\aaref@jnl{J.~Quant.~Spec.~Radiat.~Transf.}}
\def\memsai{\aaref@jnl{Mem.~Soc.~Astron.~Italiana}}
\def\nphysa{\aaref@jnl{Nucl.~Phys.~A}}   % Nuclear Physics A
\def\physrep{\aaref@jnl{Phys.~Rep.}}   % Physics Reports
\def\physscr{\aaref@jnl{Phys.~Scr}}   % Physica Scripta
\def\planss{\aaref@jnl{Planet.~Space~Sci.}}   % Planetary Space Science
\def\procspie{\aaref@jnl{Proc.~SPIE}}   % Proceedings of the SPIE

\let\amp=\&

\title[Bayesian Statistics as a New Tool for Spectral Analysis]{Bayesian Statistics
as a New Tool for Spectral Analysis: I. Application for the Determination of Basic Parameters
of Massive Stars}
\author[J-M. Mugnes and C. Robert]{J-M. Mugnes\thanks{E-mail:
jmmugnes@gmail.com} and C. Robert \\
D\'{e}partement de physique, de g\'{e}nie physique et d'optique, Universit\'{e} Laval, and
 Centre de Recherche en Astrophysique du Qu\'{e}bec,\\
 Qu\'{e}bec, Canada}
\begin{document}

\date{}

\pagerange{\pageref{firstpage}--\pageref{lastpage}} \pubyear{2015}

\maketitle

\label{firstpage}

\begin{abstract}

Spectral analysis is a powerful tool to investigate stellar properties and it has been widely used for decades now. 
However, the methods considered to perform this kind of analysis are mostly based on iteration among a few diagnostic lines 
to determine the stellar parameters. While these methods are often simple and fast, they can lead to errors 
and large uncertainties due to the required assumptions.

Here we present a method based on Bayesian statistics to find simultaneously the best combination of 
effective temperature, surface gravity, projected rotational velocity, and microturbulence velocity, using all 
the available spectral lines. Different tests are discussed to demonstrate the strength of our method, 
which we apply to 54 mid-resolution spectra of field and cluster B stars obtained at the Observatoire 
du Mont-M\'{e}gantic.  We compare our results with those found in the literature. Differences are seen 
which are well explained by the different methods used. We conclude that the B-star microturbulence velocities 
are often underestimated. We also confirm the trend that B stars in clusters are on average faster rotators than field B stars. 

\end{abstract}


\section{Introduction}

Massive stars are bright and short-lived objects. Therefore they are relatively 
easy to observe and remain close to their birth place. Their influence on their environment is important, from
photoionizing HII regions, polluting their environment with enriched chemical elements, to expelling material out of galaxies.
But they also represent an important candidate, due to their high level of internal radiation pressure, for the general study of stellar 
evolution. Therefore an accurate characterization of the fundamental and spectral parameters of massive stars is greatly needed in astrophysics.

One of the most useful tools to investigate massive stars is quantitative spectroscopy, which involves the comparison of a stellar spectrum
with synthetic data created by state-of-the-art atmospheric models. There is a wide variety of comparison technics presented in
 the literature. But whether it is by fitting spectral lines ``by eye'' \citep{Searle2008}, by using $\chi^2$ minimization 
 \citep{Daflon2007}, or by measuring equivalent widths \citep{Lefever2010}, most of these methods rely on 
an iterative process, due to the large number of parameters involved in the analysis. For B stars, these parameters are the 
effective temperature $T_{eff}$, the surface gravity $\log(g)$, the projected rotational velocity $v\sin(i)$, the micro- and 
macro-turbulence velocity $\xi$ and $\zeta$, and of course the abundances of the chemical elements. The iterative analysis usually 
starts by adopting best-estimate values for all parameters. It then alternatively refines each parameter individually while using 
specific spectral indicators that are mostly sensitive to a sub-sample of parameters. For instance, in B stars the hydrogen lines H$\beta$ 
and H$\gamma$ are mostly sensitive to $T_{eff}$ and $\log(g)$, while ionized silicon lines are mostly sensitive to $T_{eff}$, $\xi$, 
and its abundance. 
Therefore, one can start by assuming initial values for $T_{eff}$, $\log(g)$, and $\xi$ for instance, and then use the Balmer lines to determine
a new $\log(g)$ value. This new result combined with the Si lines is then used to infer a new effective temperature and, going back to the Balmer 
lines, to update $\log(g)$, and so on until all the parameter values are stable. Of course, this does not 
mean that each indicator is completely insensitive to the other parameters. On the contrary, there is a highly non-trivial coupling 
between all the involved parameters \citep{Nieva2010}. And due to this coupling, the use of an iterative process based on a few spectral 
indicators can lead to a local solution rather than a global solution that takes into account all the parameters \citep{2012ASPC..465..256M}. 
Moreover, depending on the diagnostic lines used and on the initial parameter values, the same method may not give the same results. 
Also, due to the large number of parameters involved and in order to reduce computation time, many works in the literature have focused on a 
reduced number of parameters while assigning fixed values for the others (for instance by adopting solar abundance values,
or assuming $\xi=2$~km~s$^{-1}$ for main sequence stars and $10$~km~s$^{-1}$ for evolved stars). Fixed values are often based either on 
theoretical estimates or on results from other works. For example, \citet{Searle2008} used the projected rotational velocities of
\citet{1997yCat..72840265H}. But as discussed in the present work, the final results depend on the choice of the parameter that is fixed 
and on the choice of the value adopted. 
Even for a given atmospheric model and a given set of observed spectra, all the simplifications done (i.e. the adoption of initial estimated
values, the use of fixed parameters, or an iterative process with only a few specific indicators for a reduced number of parameters) have 
an impact on the results and their accuracy and make it difficult to compare or to compile results from different works in a consistent way. 
It is therefore important to find a method which does not rely on initial estimations of the parameters, that will simultaneously 
constrain all the parameters using most of (if not all) the available information contained in an observed spectrum,
and will work in a reasonable amount of time.

Thanks to the development of computer power, along with its recent success in many scientific fields, Bayesian statistic has been more and more 
used in astrophysics during the last decades. Studies found in the literature cover a large range of astrophysical topics, from redshift 
estimations with photometry \citep{2000ApJ...536..571B} to the study of magnetic fields in interstellar clouds \citep{2010ApJ...725..466C},
including cosmological parameter selection \citep{2007MNRAS.378...72T}. Among these studies, a few 
authors applied Bayesian statistics to the determination of stellar parameters, hereby illustrating some of its advantages. For instance
\cite{2004MNRAS.351..487P} used Bayesian statistics to redefine the age-metallicity relation in field dwarfs. 
They showed that most of the scatter found in the previously derived age-metallicity relation was due to a simplified treatment
which, among other issues, did not properly take into account systematic biases. \cite{2007MNRAS.377..120S} performed a complex 
and complete study of the impact of both systematic and statistical measurement errors on stellar parameters  
and their uncertainties. More recently, \cite{2014MNRAS.443..698S} developed a method which allow the combined use of photometric, 
spectroscopic, and astrometry informations for the determination of stellar parameters and metallicities. 
Nevertheless, none of these studies were applied on massive stars, nor did they performed a spectral analysis involving more than three
parameters at the same time (the spectroscopic part of the work of Sch\"{o}nrich \& Bergemann 2014 focused only on the effective temperature, 
surface gravity, and global metallicity). 

In this work, we use both a large pre-calculated grid of synthetic spectra and small interpolated and tailored grids in order to
reduce the computation time. Bayesian statistics are applied to allow for a logical and numerical 
connection between the two grids as well as to constrain all the parameters at the same time while using nearly all the available information 
given by a spectrum. In the present paper, as a way to introduce our method and to illustrate the parameter interdependency, 
we focus on four stellar parameters (effective temperature, surface gravity, projected rotational velocity, and microturbulence velocity). 
In a following paper, we will add individual chemical abundances and the global metallicity to the analysis.

The next section summarises the ingredients and vocabulary used in Bayesian statistics. This is followed by a description
 of our technic and different tests done to demonstrate its reliability. stellar parameters are obtained for a sample of B stars and 
 are compared to other works. 

\section[]{Description of the Method}

We here briefly describe the key elements of Bayesian statistics involved in our method. 
For a complete description of Bayesian statistics see \cite{Gregory2010}. 

\subsection{The Bayes Theorem}

\subsubsection{Definition}

Let us make an experiment or an observation that allow us to retrieve some data $D$ (a stellar
spectrum for instance) in order to test a given hypothesis $H$ (for example, the hypothesis that
a given atmospheric model reproduce correctly the stellar spectrum). And let us also 
assume that we have some prior knowledge or information $I$ (the stellar parameters given by
an approximate spectral type for instance). Then by using the Bayes theorem  we can assign a
plausibility degree (or probability) to the hypothesis $H$. Namely, the Bayes theorem states that
the probability $P$ of $H$ being true given $D$ and $I$ is called the posterior probability and
is given by: 
\begin{align}
P(H | D, I)=\frac{P(H| I)*P(D|H,I)}{P(D|I)},
\end{align}
where $ P(H|I)$ is the prior probability which is obtained over a previous set of data 
or from previous knowledge, $P(D|H,I)$ is the likelihood, i.e. the result of the actual 
data $D$, and where the global likelihood $P(D|I)$ is a normalization factor. 
Simply put, according to the Bayes theorem, our final state of knowledge (the posterior 
probability) is the product of what we already knew (the prior probability) with what the 
experiment (the likelihood) tells us.

\begin{figure*}
\includegraphics[scale=0.1]{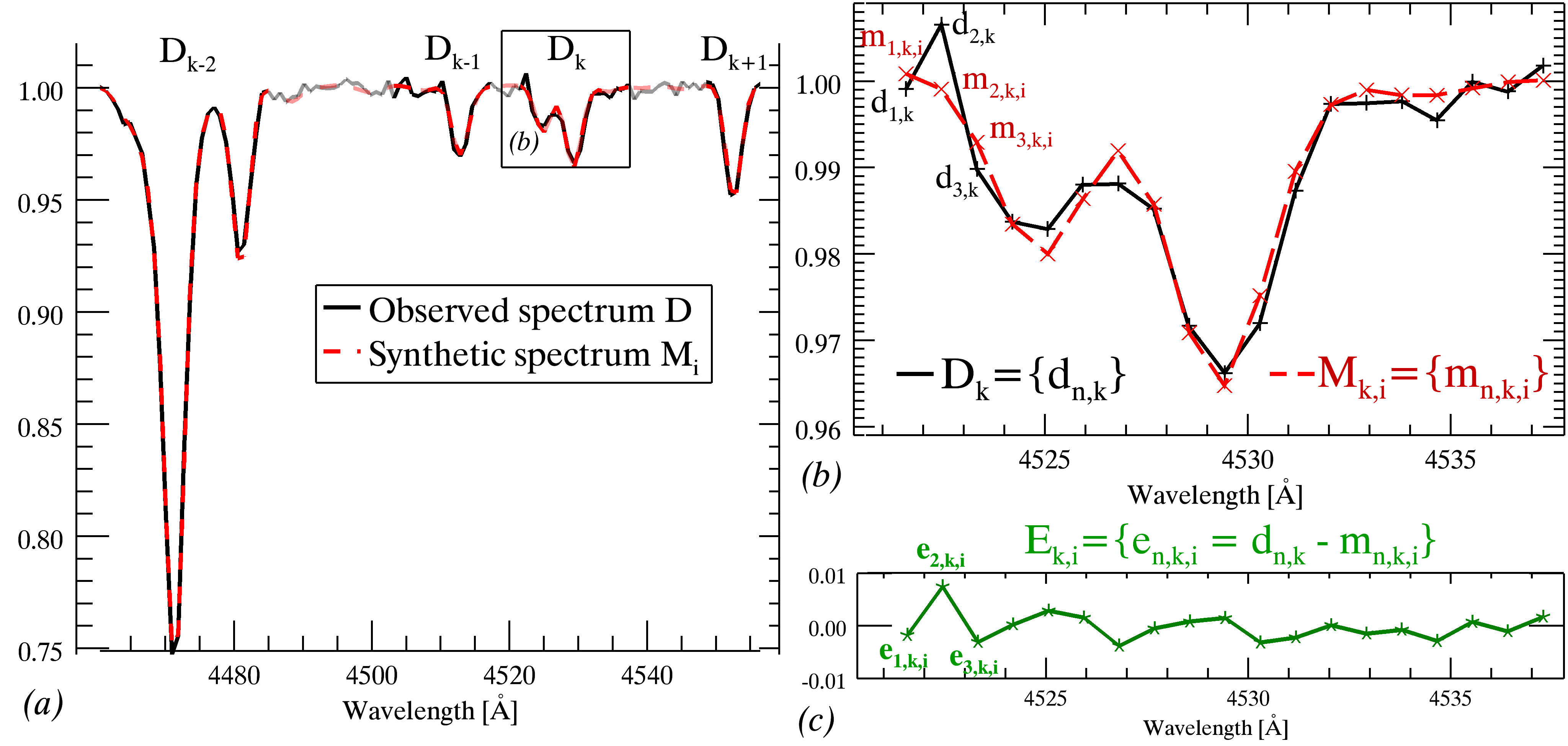}
  \caption{Illustration of the data sets. (a): Superposition of an observed spectrum $D$ (black) and synthethic spectrum $M_i$ (dashed red).
    Lighter parts of the spectra are not considered in the analysis.
    (b): Close up of one sub-spectrum $D_k$ where each observed and synthetic spectral resolution elements are labeled
    $d_{n,k}$ and $m_{n,k,i}$, respectively, with $n$ ($1 \leq n \leq N_k$) the number of resolution elements in $D_k$, 
     $k$ ($1 \leq k \leq~K$) the number of sub-spectra, and $i$ ($1 \leq i \leq Nb_{model}$) the number of synthetic spectra. 
    (c): Errors $e_{n,k,i}$, assuming that the observed spectrum is defined as $d_{n,k}=m_{n,k,i}+e_{n,k,i}$.}
\label{fig:decomp}
\end{figure*}

\subsubsection{Application to a Stellar Spectrum}

In this work we adapt the Bayes theorem to stellar spectroscopy considering three main concepts :

\noindent 1) Let us consider an observed spectrum of a given star that contains several spectral lines. We then
artificially divide the whole spectrum into smaller spectra, each centered on an individual spectral line (Fig. \ref{fig:decomp}a). 
Each smaller spectrum can be seen as an individual observation.
As the original spectrum can be described as a data set $D$ composed of all the spectral resolution elements 
(or flux points) contained in the whole spectrum, then each smaller spectrum $k$ can also be described as a data set $D_k$ 
composed of all the spectral resolution elements contained in the spectrum $k$ (Fig. \ref{fig:decomp}b). Thus for 
a given star, we have $K$ independent spectra, where each spectrum $k$ is centered on the spectral line $k$ and is described by a 
data set $D_k$ (where $1 \leqslant k \leqslant K$, and $K$ is the total number of lines contained in the original observed spectrum).

\noindent 2) As we want to fit an observed spectrum over a grid of synthetic spectra, 
we can replace the hypothesis $H$  by a set of hypothesis ${M_{i}}$ (with $i=1,2,...,Nb_{model}$) representing 
the propositions stating that each \textit{synthetic spectrum calculated with an atmosphere model $M$ with a set of parameters
$“i”$ is true} (i.e. correctly reproduces the observation). Then for a given data set $D_{k}$, 
we can assign a probability for each synthetic spectrum. 
Since the propositions ${M_{i}}$ are mutually exclusive, we can write : $ \sum_{i} P(M_{i} | D,I) =1$.

\noindent 3) Finally each line $k$ (or experiment $k$) has an associated prior information $I_k$ and a posterior probability
reflecting our state of knowledge respectively before and after doing the experiment $k$. \\

We can then rewrite the Bayes theorem for each line $k$ as :
\begin{align}\label{BT1}
P(M_{i} | D_{k}, I_{k}) =\frac{P(M_{i} | I_{k}) *P(D_{k}| M_{i}, I_{k})}{P(D_{k} | I_{k})}. 
\end{align}
If we use the posterior probability given by the line $k-1$ as the prior probability of the line $k$, then
we can define the prior information of the line $k$ with $I_k=D_{k-1},I_{k-1}$ (i.e. the combined
information given by the previous data set $D_{k-1}$, or line $k-1$, and its associated prior information $I_{k-1}$).
The Bayes theorem then becomes:
\begin{align}\label{BT}
P(M_{i} | D_{k}, I_{k}) =\frac{P(M_{i} | D_{k-1},I_{k-1}) *P(D_{k}| M_{i}, I_{k})}{P(D_{k} | I_{k})}, 
\end{align}
which can be understood as: the probability that the synthetic spectrum $M_{i}$ reproduce well the observed line $D_k$ 
is proportional to the quality of the fit (likelihood) between the observed line $D_k$ and the corresponding synthetic line from
the spectrum $M_i$ times the probability that the same synthetic spectrum $M_{i}$ reproduce the previous line $D_{k-1}$ with the previous 
corresponding synthetic line. 

Here $P(M_i | D_k, I_k)$ is the posterior probability for one synthetic spectrum $M_i$. When we consider all the synthetic spectra at the
same time, $\{P(M_i | D_k, I_k)\}_{i=1..Nb_{models}}$ is thus called the posterior probability distribution (the same obviously applies for the 
likelihood and the prior).

In other words, if we apply this theorem to all the lines in the observed spectrum (each time using the last posterior 
probability distribution as a prior for the next line) and considering all the synthetic spectra, we obtain, in the end, 
a final posterior probability distribution that simultaneously takes into account all the constraints given by all the lines in the 
observed spectrum for all the considered parameters. Figure \ref{fig:methex}, shows an example of the process when we apply
our method on 4 spectral lines using 25 synthetic spectra.
Note here that the order of the lines to which we apply the theorem does not matter. Indeed, if we neglect the normalisation factors, 
the final posterior probability is simply the product of the first prior probability with the likelihoods of all the lines involved.

\subsubsection{Constructing the Likelihood}\label{sigm}

In order to construct a proper likelihood, we need to make a few basic assumptions. First, each 
data set $D_k$ (i.e. line $k$) is composed of $N_k$ independent spectral resolution elements
$D_k=\{d_{n,k}\}= d_{1,k}, d_{2,k},...,d_{N_k,k}$ (Fig. \ref{fig:decomp}b). Then we can do the same for the model spectrum 
$M_{i,k}=\{m_{n,k,i}\}= m_{1,k,i}, m_{2,k,i},...,m_{N_k,k,i}$ (Fig. \ref{fig:decomp}b). Finally let us suppose that each observed 
spectral resolution element can be associated with a spectral resolution element from a model, given a certain flux 
error $d_{n,k}=m_{n,k,i}+e_{n,k,i}$ (Fig. \ref{fig:decomp}c). If we suppose the errors $\{e_{n,k,i}\}$ to be independent and 
 represented by a Gaussian distribution centered on zero with a standard deviation of $\sigma_{n,k}$, then we can write the likelihood as:

{
\begin{align}\label{Lik}
P(D_{k} | M_{i},I_{k}) & =\prod_{n=1}^{N} \left[ \frac{1}{(\sqrt{2\pi}\sigma_{n,k}} \right]* \notag\\ 
  &  \exp\left[-\sum_{n}\frac{(d_{n,k}-m_{n,k,i})^{2}}{2\sigma_{n,k}^{2}} \right]   \notag\\
& = C_{k} \exp\left[-\frac{\chi_{i,k}^{2}}{2}\right].
\end{align}
}%

The errors can represent the noise in the data but also the incompleteness of the model (due to simplifications, 
approximations, or unknowed variables) or even some errors in the reduction process of the data (bad normalisation 
for instance). To take these sources into account, we define the standard deviation of the gaussian distribution as follow:
{
\begin{align}\label{sigmas}
\sigma_{n,k}=\sqrt{\smash[b]{(\sigma_{noise})_{n,k}^2+(\sigma_{model})_{n,k}^2}},
\end{align}
}%
where $(\sigma_{noise})_{n,k}$ refers to the errors coming from the noise in the data, and $(\sigma_{model})_{n,k}$ refers to 
the other sources of errors (model incompleteness and reduction errors). In this work, for simplicity, we assume  
$(\sigma_{noise})_{n,k}=\sigma_{noise}$, a constant for all the spectrum, defined as the standard deviation in regions of the normalised 
continuum of the observed spectrum. And for each line, $(\sigma_{model})_{n,k}$ is given by:
{
\begin{align}\label{sigma_mod}
(\sigma_{model})_{n,k} & =\min_{\forall i} \left( \sqrt{ \smash[b]{\frac{1}{N_k}\sum_{n}( d_{n,k}-m_{n,k,i} )^{2} }} \right) \notag\\ 
 & =(\sigma_{model})_{k}.
\end{align}
}%
In other words, for each line $k$, $(\sigma_{model})_{k}$ is the minimal RMS between the observed line and all the corresponding
model spectrum $M_{i,k}$. %(ou ``is the RMS residual of the best model spectrum $M_{i,k}$.'')
With these definitions, $\sigma_{n,k}$ is the same for all the spectral resolution elements $n$ of a line, but is different for each line.
Thus: 
{
\begin{align}\label{sigma_chi}
\sigma_{n,k} & =\sqrt{\smash[b]{\sigma_{noise}^2+(\sigma_{model})_{k}^2}} = (\sigma_{\chi})_{k}.
\end{align}
}%
Note that $(\sigma_{\chi})_{k}$ is a very important variable as it ultimately links physical imperfections and theoretical incompleteness
to the uncertainty of the final parameters. And with this decomposition, if we were to work with perfect data ($\sigma_{noise}\longmapsto0$), 
then the uncertainty of the final parameters would be related to the limitations of the atmospheric or atomic models used.
Inversely, if we have a perfect model ($(\sigma_{model})_{k}\longmapsto0$), the final uncertainties would then be directly related to the 
quality of the data. 

\subsubsection{Global Likelihood and First Prior}

Now, since $\sum_i P(M_i | D_k,I_k) =1$, the global likelihood $P(D_k|I_k)$ is then given by: 
\begin{align}
P(D_k | I_k)= \sum\limits_i \left( P(M_i | D_{k-1},I_{k-1}) *P(D_k | M_i, I_k) \right).
\end{align}
And the Bayes theorem (Eq. \ref{BT}) becomes:
\begin{align}\label{FBT}
P(M_i | D_k, I_k) = \frac{P(M_i | D_{k-1},I_{k-1}) *C_{k} \exp\left[-\frac{\chi_{i,k}^{2}}{2}\right]}{ \displaystyle\sum\limits_i \left( P(M_i | D_{k-1},I_{k-1}) * C_{k} \exp\left[-\frac{\chi_{i,k}^{2}}{2}\right]\right)}. 
\end{align}

\begin{figure*}
\includegraphics[scale=0.12]{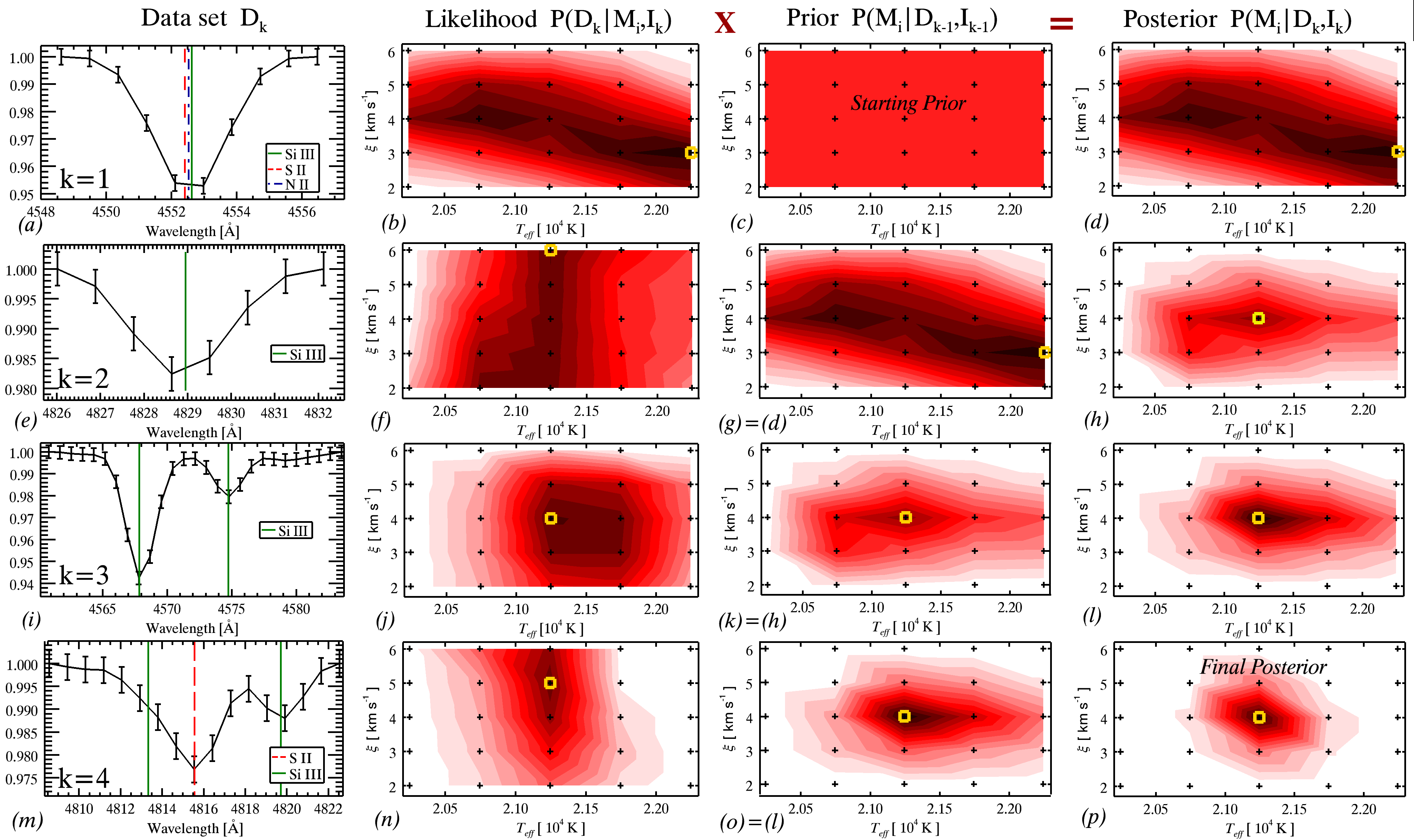}
\caption{ 
Illustration of the Bayesian analysis step by step. For this example, we consider 4 data sets (or lines) $D_k$ and 25 synthetic spectra calculated 
with only two free parameters: $T_{eff}$ and $\xi$. The black plus signs in columns 2, 3, and 4 indicate the value used for these
two parameters. The colour of the surface contours is scaled to the relative probability of each distribution: the dark red regions being 
the most probable and the white ones the less probable. The yellow squares indicate the global maximum of each distribution.  
The first column shows every data set considered, and the second column the likelihood obtained for each data set as in Equation \ref{Lik}.
The third column represents the prior probability distribution known at each step $k$, and the last column shows the posterior probability 
distribution which is the normalised product of the prior and the likelihood of a given step $k$. The method goes as follow: considering the 
first line $k=1$ (in (a)), we calculate its likelihood using Equation \ref{Lik} (in (b)), and use, as a first prior, a flat distribution 
(in (c)) assuming we have no knowledge of the best pair [$T_{eff}$, $\xi$]. We then apply the Bayes theorem (Eq. \ref{BT}) and obtain the 
first posterior probability distribution (in (d)). For the next data set ($k=2$), we perform the same operation but we use the posterior
distribution of the previous step (in (d)) as a new prior distribution (in (g)) since our ``state of knowledge'' has been changed 
after the first step. In the end, the final posterior probability distribution (in (p)) represents the solution combining all the 
information given by the 4 lines. Note that since the method can be summarized as a series of products, the order in which we
consider the data sets have no influence on the shape of the final posterior distribution (see Fig. \ref{fig:comex}).}
\label{fig:methex}
\end{figure*}

Finally, when we apply the theorem (Eq. \ref{FBT}) for the first time, we assume that all models $M_i$ are equiprobable:
$P(M_i| D_0,I_0)= 1/Nb_{model}$ (i.e. we consider a flat prior distribution as in Fig. \ref{fig:methex}c, 
and $D_0$ represents no data). This means that, at first, we do not give preference to any model (or set of parameters). 
Note here that we are not performing a model comparison but a parameter estimation and as we use large data samples, 
the results are not affected by the choice of the prior \citep{Gregory2010}.

\subsection{Parameter Estimation}\label{param_estim}

In this work we focus on the stellar parameters $T_{eff}$ the effective temperature,
$\log(g)$ the surface gravity, $v\,\sin(i)$ the projected rotational velocity, and $\xi$ the microturbulence velocity.
We thus need to retrieve, from the final posterior probability mentioned above, an estimation for each parameter by 
considering all the other parameters. For this, we use another major feature of the Bayesian statistics, the marginalization. 
This is done by defining the model $M_i$ as: \textit{the synthetic spectrum computed with the model $M$
using the set of parameters $T_{effi}$, $\log(g)_i$, $v\,\sin(i)_i$, and $\xi_i$}, which we write $Mi=T_i,l_i,v_i,\xi_i,M$.
Since we suppose our model $M$ to be true, our posterior probability thus becomes:
\begin{align}
P(M_i | D_k, I_k)= P(T_i,l_i,v_i,\xi_i | M,D_k,I_k),
\end{align}
which can be read as the joint probability for $T_i$, $l_i$, $v_i$, and $\xi_i$ to be true if the model $M$, the data $D_k$,
and our prior knowledge $I_k$ are true.

\begin{figure}
\includegraphics[scale=0.06]{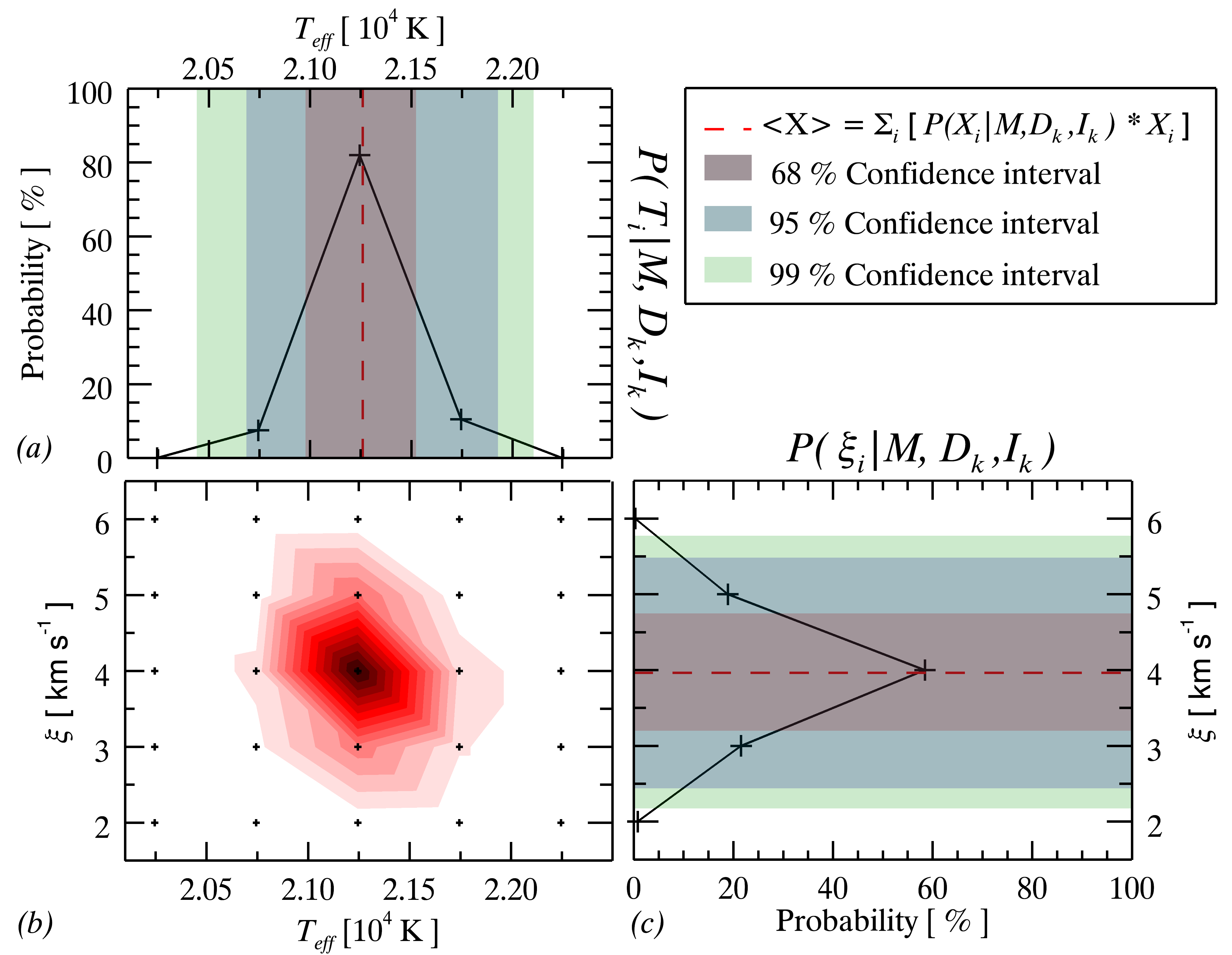}
  \caption{
  Exemples of the marginal probability distribution for $T_{eff}$ and $\xi$. These distributions (in (a) and (c)) 
  are derived from the final posterior probability distribution (in (b)) taken from Figure \ref{fig:methex}p. The coloured regions
  in (a) and (c) indicate the $68\%$, $95\%$, and $99\%$ confidence intervals which can be seen as a $1\sigma$, $2\sigma$,
  and $3\sigma$ uncertainties for the best value of each parameter. The dashed red lines show the position of the mean value of each
  parameter defined as $\langle X \rangle = \sum_i  X_i * P(X_i | M,D_k,I_k)$, where $X$ is one of the considered parameters. 
  Since the parameter space is discrete, the most probable value found for a parameter might not be the true value but the closest 
  value used in the calculation of the spectra. Thus the mean value can be a better estimation of the real value of a parameter. 
}
\label{fig:margex}
\end{figure}

Since our parameters have discrete values in our analysis, we can retrieve the marginal posterior
probability distribution for each parameter by using the marginalization equation. Here we
give an example for $T_i$ :
\begin{align}
P(T_i |M,D_k,I_k)= \sum_a \sum_b \sum_c P(T_i,l_a,v_b,\xi_c|M,D_k,I_k),
\end{align}
which is the marginal probability distribution of $T_i$ considering all the values of the other parameters. 
As shown in Figure \ref{fig:margex}, from this distribution we obtain the most probable value and the mean value, 
as well as the associated uncertainties which are defined as the values between which the sum of the marginal posterior 
probability distribution is equal to or greater than a certain confidence level $C=\sum_{T_-}^{T_+} P(T_i | M,D_K,I_K)$. 
In this work we use $C=0.99$ which can be seen as a $3\sigma$ uncertainty (recall that $\sum_i  P(T_i | M,D_K,I_K)=1$).
Note that these uncertainties are related to the variations of the other parameters but also, as stated in $\S\,$\ref{sigm},
to possible data or model imperfections.

\subsection{Preventing Numerical Biases}

Due to the fact that we want to treat all available lines equally, we need to calculate the $\chi^2$
for the same number of spectral resolution elements in each line. Otherwise, broad features (such as hydrogen 
and helium lines in B stars) will dominate the solution over the narrow features (metal lines for instance) in the posterior probability. 

Indeed, given the same differences $d_{n,k}-m_{n,k,i}$ between the observed and synthetic line and the same $\sigma_{\chi}$
for both a broad and a narrow line, only the number of spectral resolution elements will have an impact on the $\chi^2$ value. 
The more spectral resolution elements there are, the higher is the value of the $\chi^2$. Therefore, the $\chi^2$ of a broad line will 
rapidly increase and hence its likelihood will also rapidly tend to zero as the models differ from the solution, whereas a narrow line
will see its likelihood slowly reach zero. Consequently when multiplying the prior probability density given by the broad 
line with the likelihood of the narrow line, only the broad line solution survives. To prevent this effect, we therefore interpolate 
each line with the same number of flux points given by the broadest line in the sample. 
This interpolation will obviously oversample the narrow lines, but it is a good way to consider each line equally without
losing information (since interpolating each line with the number of flux points given by the narrowest line would imply
a loss of detail in the broader lines).
The oversampling has also the advantage of giving a more realistic weight to the narrow and weaker lines; the parameter space 
for the solution of a weaker line is then better defined. 

\subsection{Synthetic Spectra}\label{method}

We use atmospheric models from the metal line-blanketed, NLTE, plane-parallel, and hydrostatic
code TLUSTY \citep{Lanz2007}. With these models, we create synthetic spectra by running the program SYNSPEC. More precisely, we use
the SYNPLOT package generously offered to us by Hubeny and Lanz, which contains a TLUSTY pre-calculated B-star grid (\textit{BSTAR2006}) and
the latest version of SYNSPEC and SYNPLOT which allows the user to interpolate more specific models from the closest \textit{BSTAR2006} models.
The \textit{BSTAR2006} grid considers 16 values of the effective temperature, between $15000$~K and $30000$~K with a $1000$~K step; 
13 surface gravities, from $1.75$ to $4.75$ in $\log(g)$ with a step of $0.25$ dex; and 6 chemical compositions, $Z/Z_{\sun}=2,1,1/2,1/5$, 
and $1/10$ (where $Z_{\sun}$ is the solar composition); a solar helium abundance, He/H = 0.1 by number, and a microturbulence velocity of
$2$ and $10$~km~s$^{-1}$. 

For this paper we consider only the solar chemical composition, as established by \citet{Asplund2009}, and we use SYNSPEC to create a 
grid of synthetic spectra (hereafter called \textit{BaseGrid}) with the same values of the effective temperature and surface gravity as 
those of the \textit{BSTAR2006} grid with an extended range of $7$ microturbulence velocities, from $0$ to $30$~km~s$^{-1}$ with a step of 
$5$~km~s$^{-1}$. We then convolve these spectra to take into account rotational broadening with 21 different values of $v\,\sin(i)$, 
from $0$ to $400$~km~s$^{-1}$ with a $20$~km~s$^{-1}$ step, to create our starting grid for the analysis: \textit{BSTART}. Note 
that the spectra of \textit{BaseGrid} are not convolved with instrumental and rotational broadening and are used, in this work, as the 
basis for the creation of a new spectrum or new refined grids by directly interpolating the non-convolved spectra (because this is faster 
than interpolating the \textit{BSTAR2006} models) to the desired values of $T_{eff}$, $\log(g)$, and $\xi$.

For each star, we begin, as previously mentioned, by assuming a uniform prior probability over all the \textit{BSTART} models.
We then apply our Bayesian analysis over all the selected lines in the spectrum of a star using all the models of the \textit{BSTART} grid. 
This first analysis gives us, after marginalization, a reduced parameter space by excluding all values outside the $99\%$ confidence level.
We then construct a refined grid in this new parameter space and we reanalyze the lines assuming,  
once more, a flat prior. The result gives an even more restricted parameter space over which we reconstruct another refined grid
and redo the analysis, and so on. The program stops either when the parameter space remains unchanged or 
when the parameter steps used to create a new grid are equal to or smaller than a certain limit. Here, 
as a compromise between accuracy and computing time, we set this limit at $100$~K in $T_{eff}$, $0.03$ dex
in $\log(g)$, $5$~km~s$^{-1}$ in $v\,\sin(i)$, and $1$~km~s$^{-1}$ in $\xi$.

\section[]{Testing the Method on Artificial Stellar Spectra}

As a way to test the efficiency of the method as well as its robustness to noise and uncertainties, we perform three tests.

\begin{figure}
\includegraphics[scale=0.26]{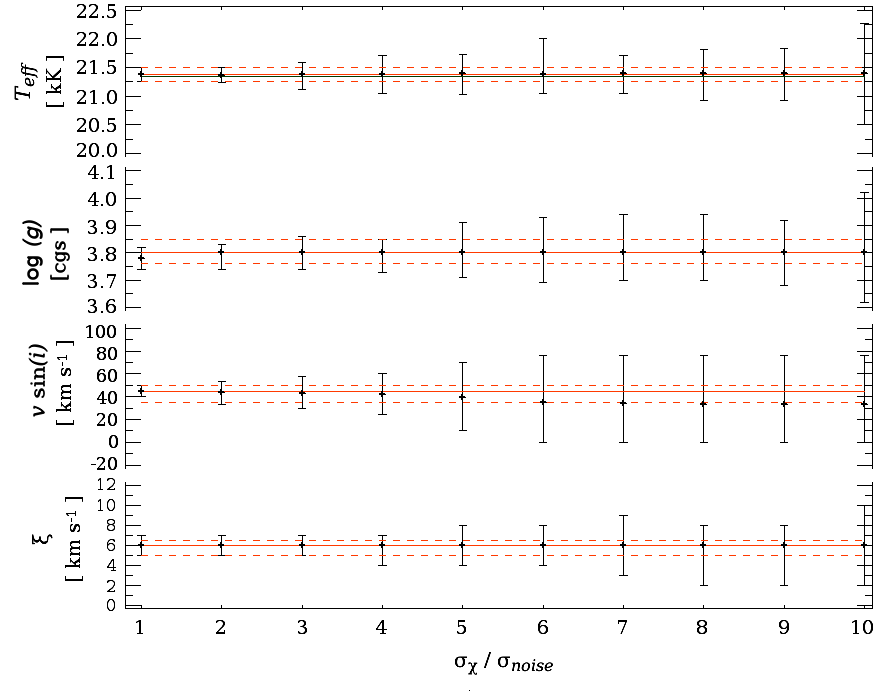}
  \caption{Plot of the stellar parameter values for 10 runs with increasing $\sigma_{\chi}/ \sigma_{noise}$. 
Individual points and error bars are the results of the different runs.
The green lines show the real parameter values used to create the artificial stellar spectrum ($T_{eff}=21350$~K,
$\log(g)=3.80$, $v\,\sin(i)=45$~km~s$^{-1}$, and $\xi=6$~km~s$^{-1}$).
The orange lines show the results (full lines) and the corresponding uncertainties (dashed lines) from the analysis of
the artificial spectrum without noise added and for $\sigma_{\chi}/\sigma_{noise} =1$.
The orange lines often overlap with the green lines.}
\label{fig:msig}
\end{figure}

For the first test, we create an artificial observational spectrum using synthetic spectra with the following specifications:
$T_{eff}=21350$~K, $\log(g)=3.80$, $v\,\sin(i)=45$~km~s$^{-1}$, $\xi=6$~km~s$^{-1}$, and solar chemical composition. 
The spectrum is also convolved with a Gaussian instrumental profile with a full width at half maximum (FWHM) of $2.3$~\AA{}.
Random noise is also added in order to emulate a standard deviation, $\sigma_{noise}$,
that is typical of what we see in our actual observations (with $\sigma_{noise}$ $\simeq$ 0.004 in the normalised flux). Then, we 
compare the artificial spectrum with our models using the method described in $\S\,$\ref{method}. But, as we are in the situation 
of a perfect model and imperfect data, we set $(\sigma_{model})_{k}$ to 0 and we substitute $\sigma_{noise}$ by $x\sigma_{noise}$ 
where $x=1,2,3,...$ in Equation \ref{sigma_chi}. This way, while not changing the actual noise level in the spectrum, we can overestimate 
its value in the calculation by increasing $x$, and thus, we can test the robustness of the method to noise over-estimation. 
This analysis is then repeated 10 times, each time increasing $x$ by $1$. The results are summarized in Figure \ref{fig:msig}.
 
While running the analysis on the artificial spectrum without noise added (orange lines in Fig. \ref{fig:msig}), the program returns
almost the exact parameter values used to create this artificial spectrum (green lines in Fig. \ref{fig:msig}). There is only a slight
difference of $10$~K for the effective temperature which is well under our minimal uncertainty ($100$~K). The runs for different values
of $\sigma_{\chi}/ \sigma_{noise}$ then give parameter values which are all very close to the original values used to create the 
artificial spectrum.

\begin{figure}
\includegraphics[scale=0.32]{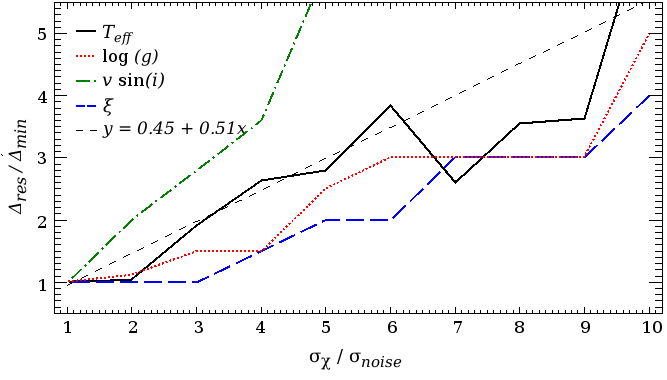}
  \caption{Plot of the relative stellar parameter uncertainties $\Delta_{res}/\Delta_{min}$ with increasing $\sigma_{\chi}/\sigma_{noise}$.
For each parameter, $\Delta_{res}$ is the length of the error bars from Figure \ref{fig:msig} and $\Delta_{min}$ the length of 
the smallest error bar found for the 10 runs. The thin black dashed line is a linear fit of the mean of all
the parameter uncertainties.}
\label{fig:statchi}
\end{figure}

Increasing the ratio $\sigma_{\chi}/\sigma_{noise}$ means overestimating the noise level in the artificial spectrum, which
should logically lead to greater parameter uncertainties since $\sigma_{\chi}$ acts as a tolerance parameter. Here, the parameter
uncertainties increase with $\sigma_{\chi}$ roughly as $\sigma_{noise}/2$, as shown in Figure~\ref{fig:statchi}. This means that even 
when increasing the flux uncertainty by a factor of $2$ or $3$, there will be nearly no incidence on the method accuracy. Note also that,
up to $4\sigma_{noise}$, the accuracy of $v\,\sin(i)$ is largely higher than the adopted resolution 
(around $77$~km~s$^{-1}$ for a FWHM of $2.3$~\r{A}). Usually, with classical methods, the uncertainty for $v\,\sin(i)$ with a value smaller 
than half the FWHM should be of the order of FWHM/2 since instrumental broadening becomes more important than rotational broadening under 
this limit. Theses effects are a direct consequence of cumulating the information from one line to the others and thus having 
multiple simultaneous constraints on each parameter. Note that when we use large $\sigma_{\chi}$ values, we find the classical
uncertainties (close to the $v\,\sin(i)$ value when $v\,\sin(i)$ is below the classical limit) since, with such a huge tolerance 
parameter, most of the constraints coming from the lines disappears.    

\begin{figure}
\includegraphics[scale=0.32]{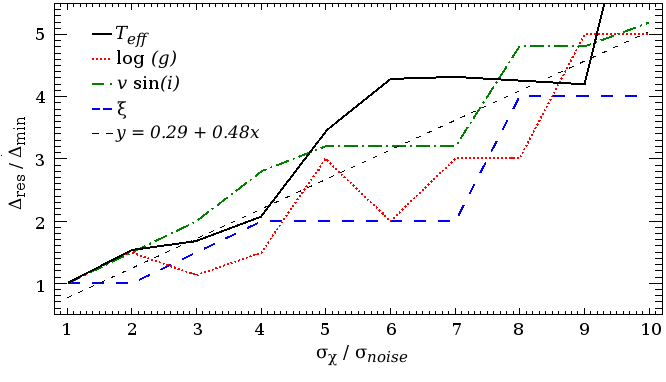}
  \caption{Plot of the relative stellar parameter uncertainties $\Delta_{res}/\Delta_{min}$ with 
increasing $\sigma_{\chi}/\sigma_{noise}$ for a higher projected rotational velocity, $v\,\sin(i)=145$~km~s$^{-1}$. 
Uncertainty and line definitions are the same as in Figure \ref{fig:statchi}.}
\label{fig:statchivr}
\end{figure}

Since spectral analysis usually tends to be more difficult and less accurate for fast rotating stars due to the blending 
of multiple lines, we repeat the first test for a higher projected rotational velocity, $v\,\sin(i)=145$~km~s$^{-1}$. All
the other parameters are unchanged (we are using the same lines as well for this analysis).
The evolution of the parameter uncertainties is shown in Figure \ref{fig:statchivr}. 
Here, the effect seen in Figures \ref{fig:msig} and \ref{fig:statchi} for a low $v\,\sin(i)$ value (compared to the instrumental FWHM)
disappears while the slope of the fit shown in Figure \ref{fig:statchivr} remains the same as in Figure \ref{fig:statchi}. Also, the 
uncertainties of each parameter are of the same order as those obtained for a slow rotator. As this result is the same for an 
even faster rotator (for instance $v\,\sin(i)= 300$~km~s$^{-1}$), we can conclude that a faster rotational velocity does not affect the accuracy 
on the other stellar parameters when using our method. 
Again this is a consequence of cumulating the information from one line to the others.

\begin{figure}
\includegraphics[scale=0.30]{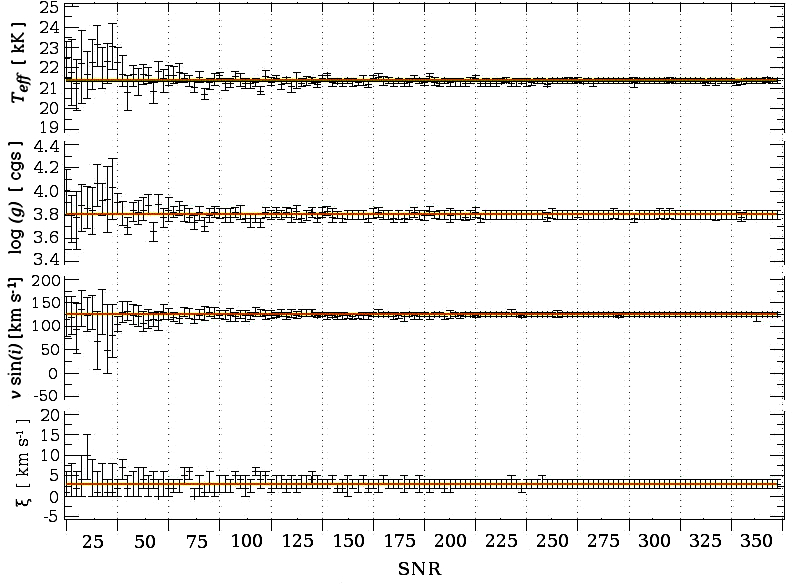}
  \caption{Plot of the stellar parameter values as a function of the signal-to-noise ratio. For each SNR value,
10 artificial spectra with a random noise structure are tested.
The green lines show the real parameter values used to create the original artificial spectrum ($T_{eff}=21350$~K,
$\log(g)=3.80$, $v\,\sin(i)=125$~km~s$^{-1}$, and $\xi=3$~km~s$^{-1}$).
The orange lines show the best parameters returned for the artificial spectrum
without noise added.
The orange lines often overlap with the green lines.}
\label{fig:snrsig}
\end{figure}

\begin{figure}
\includegraphics[scale=0.34]{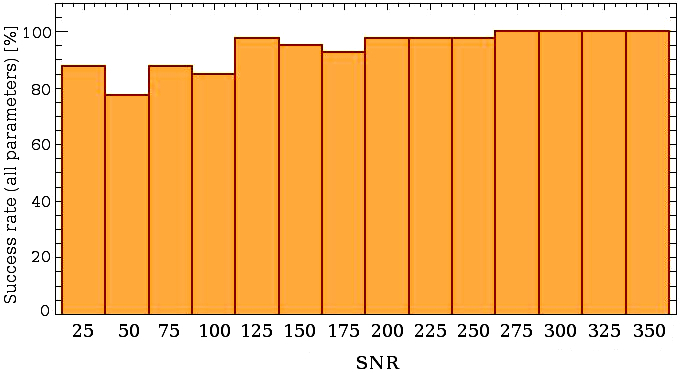}
  \caption{Success rate of the method, to derive the correct parameter values (within the uncertainties) used to create the 
  original artificial spectrum, as a function of the signal-to-noise ratio. 
  Percentages are mean values gathered from all the parameters for each SNR value.}
\label{fig:pourcsig}
\end{figure}

For the second test, we again use an artificial spectrum (with the following parameters:
$T_{eff}=21350$~K, $\log(g)=3.80$, $v\,\sin(i)=125$~km~s$^{-1}$, $\xi=3$~km~s$^{-1}$, and solar chemical composition)
but this time we change the noise level added to it in order to create different signal-to-noise ratios (SNR), 
ranging from 25 to 350 per resolution element with a step of 25. We estimate the SNR using the 
\textit{der\_snr} program of \citet{2008ASPC..394..505S}. On this scale, a SNR of 250 corresponds to a $\sigma_{noise}$
of approximately $0.004$ in the normalized continuum. For each value of SNR, we create 10 noisy artificial spectra.
For the 140 new spectra thus created, we apply our method with the original definition of $\sigma_{\chi}$ (Eq. \ref{sigma_chi}).
This way we simulate a real analysis even though we still are in the case of a perfect model and imperfect data. Of course,
in this case $\sigma_{\chi}$ is overestimated by roughly a factor of 2, since for a given line $k$, $(\sigma_{model})_{k}\approx\sigma_{noise}$,
but as explained above, it has nearly no incidence on the method accuracy. The final parameters returned by the 140 runs are shown in 
Figure \ref{fig:snrsig} and the overall success rate of the method is shown in Figure \ref{fig:pourcsig}. This test allow us to 
check for the robustness of the method against highly noisy data, and we can see that for SNR around 100-125 and higher,
the method returns correct and accurate results (with uncertainties equal or close to the minimal uncertainties and with a mean 
success rate of $97.5\%$). For lower SNR ($\leqslant100$), the parameter uncertainties become naturally larger but the real 
parameter values are still found nearly $84\%$ of the time (within the uncertainties). Considering
all the SNR values, the overall success rate is around $94\%$. 

\begin{figure}
\includegraphics[scale=0.30]{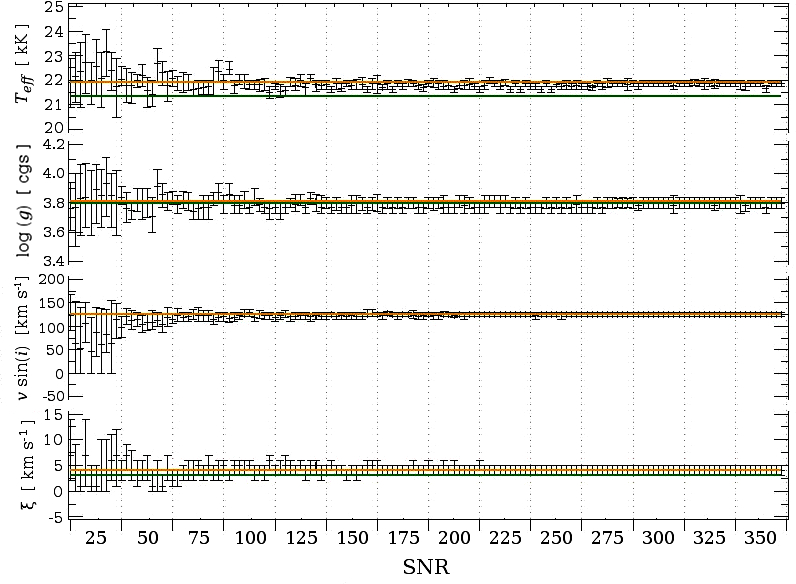}
  \caption{Same test as in Figure \ref{fig:snrsig} when using an artificial spectrum created by the atmospheric model ATLAS9.
The green lines show the real parameter values used to create the original artificial spectrum ($T_{eff}=21350$~K,
$\log(g)=3.80$, $v\,\sin(i)=125$~km~s$^{-1}$, and $\xi=3$~km~s$^{-1}$).
The orange lines show the best parameters returned for the original spectrum without noise added ($T_{eff}=21930$~K,
$\log(g)=3.81$, $v\,\sin(i)=125$~km~s$^{-1}$, and $\xi=4$~km~s$^{-1}$).
The orange line overlaps the green line in the case of $v\,\sin(i)$.}
\label{fig:snrsigbk}
\end{figure}

\begin{figure}
\includegraphics[scale=0.32]{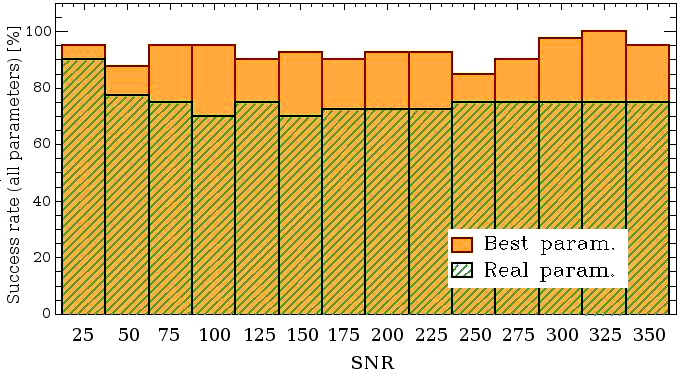}
  \caption{Success rate of the method, to derive the correct parameter values (within the uncertainties),
  as a function of the signal-to-noise ratio. Here the correct parameter values are either the values used to create the 
  hybrid artificial spectrum (\textit{Real param}), or the best values found when analysing the artificial spectrum
  without noise (\textit{Best param}).
  Percentages are mean values gathered from all the parameters for each SNR value.}
\label{fig:pourcsigbk}
\end{figure}

The third test is the same as the second one, except that the artificial spectrum is created with the atmospheric model ATLAS9 
\citep{Castelli2004}, rather than with TLUSTY, using the same stellar parameters. 
This test is close to an analysis of real data since we are in the case of imperfect data and inexact models.
The results are shown in Figure~\ref{fig:snrsigbk}, and the overall success rate of the method in 
Figure~\ref{fig:pourcsigbk}. Compared to Figure~\ref{fig:snrsig}, we can see from Figure~\ref{fig:snrsigbk} that the best parameters 
returned by the method for the spectrum without noise (orange lines) are different than the real parameters (green lines) used to create 
the spectrum. It is especially true for the effective temperature (the other parameters are within the minimal uncertainties), where 
$T_{eff Best}=21930$~K while $T_{eff Real}=21350$~K. This discrepancy is to be expected when comparing full non-LTE 
(i.e. non-LTE atmospheric models with non-LTE radiative transfert) spectra with hybrid LTE/non-LTE
(i.e. LTE atmospheric models with non-LTE radiative transfert) spectra since, non-LTE line profiles tend to be slightly 
stronger and broader than those derived from LTE atmospheres, thus giving a lower temperature when considering
a LTE treatment \citep{Huang2010,Lanz2007}. Here the artificial spectrum is created by a hybrid model while the 
grid of synthetic spectra are calculated using a full non-LTE treatment. Therefore our method finds a higher effective temperature
than the one used to create the hybrid spectrum. Note that this discrepancy is not taken into account by $\sigma_{model}$ since
this parameter reflects the model incapacity to correctly reproduce the lines (whatever parameter values are considered). 
Here, our model is able to reproduce nearly perfectly most of the lines but for a different effective temperature than the one used 
to create the hybrid lines. With this in mind, if we consider the best parameter values found for the hybrid spectrum 
without noise as the reference values (rather than the real values used to create the hybrid spectrum), we find a mean success rate of 
nearly $93\%$ (Fig. \ref{fig:pourcsigbk}) which is equivalent to the success rate found in the previous test (Fig. \ref{fig:pourcsig}). 
Note that if we consider the original parameter values, the mean success rate for all the parameters is around $75\%$, and 
around $96\%$ when we exclude $T_{eff}$. 
In tests 2 and 3, if we multiply the derived parameter uncertainties by a factor $1.5$ or $2$, the overall success 
rate (which is around $94\%$ in both case) jumps to nearly $98\%$  and $99\%$, respectively
Thus, in terms of classical statistics, our results are given with an accuracy close to a $2\sigma$ uncertainty.

\section[]{Testing our Method on Real Stellar Spectra}

\subsection{Observational Data}\label{data}

The selection criteria for our sample were as follow: 1) only mid-B to early-B type stars, since the pre-calculated \textit{BSTAR2006}
atmospheric models only cover the range of effective temperature for these subtypes; 2) relatively bright stars to allow a high SNR;
3) no binaries or peculiar stars to avoid complications while testing our method; 4) relatively close objects in order to work
with a solar chemical abundance approximation (chemical element abundances will be the topic of a following paper);
and 5) well known objects to allow the comparison of our method with other works. The main studies used here for the comparison are from
 \citet[hereafter HG10]{Huang2010}, \citet{Lefever2010}, \citet{Takeda2010}, \citet{Searle2008}, \citet{Markova2008}, \citet{Daflon2007},
 \citet[hereafter HG06]{Huang2006a,Huang2006b}, \citet{Lyubimkov2005}, and \citet{Andrievsky1999}.

We test our method on a sample of 54 mid-B to early-B stars, where 38 are from the field and 16 from clusters.
The field stars are all nearby dwarf and giant B stars, with a visible magnitude $3.8 \leq V \leq 9.7$.
The cluster stars are members of two open clusters: $8$ B stars from NGC1960 (M36) with $8.8 \leq V \leq 10.8$, and $8$ B stars
from NGC884 ($\chi$ Persei) with $8.9 \leq V \leq 11.6$. These two clusters have a colour excess $E(B-V)=0.22$ and $0.56$,
respectively, and due to their proximity, are assumed to have a metallicity close to solar \citep{Kharchenko2005}. 
The spectra have been collected at the 1.6 m telescope of the Observatoire du Mont-M\'{e}gantic at 3 epochs: February 2011, August
2013, and December 2013. They cover the wavelength range from 3500 to 5500~\r{A} with a spectral resolution of 2.3~\r{A} and a mean
SNR of 250 per pixel in the continuum (with very little variation from the blue to the red end).
We select all the observed lines available for the spectral analysis, except when they are not reproduced by the models, when 
they are barely distinct from the noise, or when they are superimposed on bad CCD pixels.
Table \ref{tab:linelist} lists all the major useful lines. Note that not all these lines are available 
 for every stars since they may appear or disappear depending on the spectral type. Note also that due to our modest 
resolution, and sometimes due to strong rotational velocities, more minor lines are included in the analysis.

\begin{table}
\caption{Principal spectral lines used for the analysis}
\begin{tabular}{|c|l|}
\hline
Species & Wavelength (\r{A}) \\ \hline
H I & 3712, 3722, 3734, 3750, 3771, 3798, 3835, 3889,\\
    & 3970, 4102, 4340, 4861 \\ \hline
He I & 3785, 3820, 3867, 3872, 3927, 3936, 4009, 4026,\\
     & 4121, 4144, 4169, 4388, 4438, 4471, 4713, 4922,\\
     & 5016, 5048 \\ \hline
C II & 4267, 4374, 4619, 4619, 5133, 5145, 5648 \\ \hline
N II & 3995, 4044, 4228, 4237, 4447, 4601, 4607, 4631,\\
     & 4803, 4994, 5667, 5680 \\ \hline
O II & 3912, 3954, 3982, 4070, 4079, 4085, 4153, 4185,\\
     & 4277, 4304, 4317, 4320, 4367, 4415, 4501, 4591,\\
     & 4642, 4649, 4662, 4676, 4705, 4907, 4943 \\ \hline
Mg II & 4481 \\ \hline
Al III & 4513, 4529, 5697 \\ \hline
Si II & 3856, 3863, 4131, 5056 \\ \hline
Si III & 4553, 4568, 4575, 4683, 4829, 5740 \\ \hline
Si IV & 4089 \\ \hline
S II & 4294, 4525, 4816, 5032, 5103, 5321, 5433, 5454 \\ \hline
S III & 4254, 4285, 4362 \\ \hline
Fe II & 5169, 5260 \\ \hline
Fe III & 5074, 5087 \\ \hline
\end{tabular}
\label{tab:linelist}
\end{table}

\subsection{Results and Discussions}

\begin{figure*}
\includegraphics[scale=0.12]{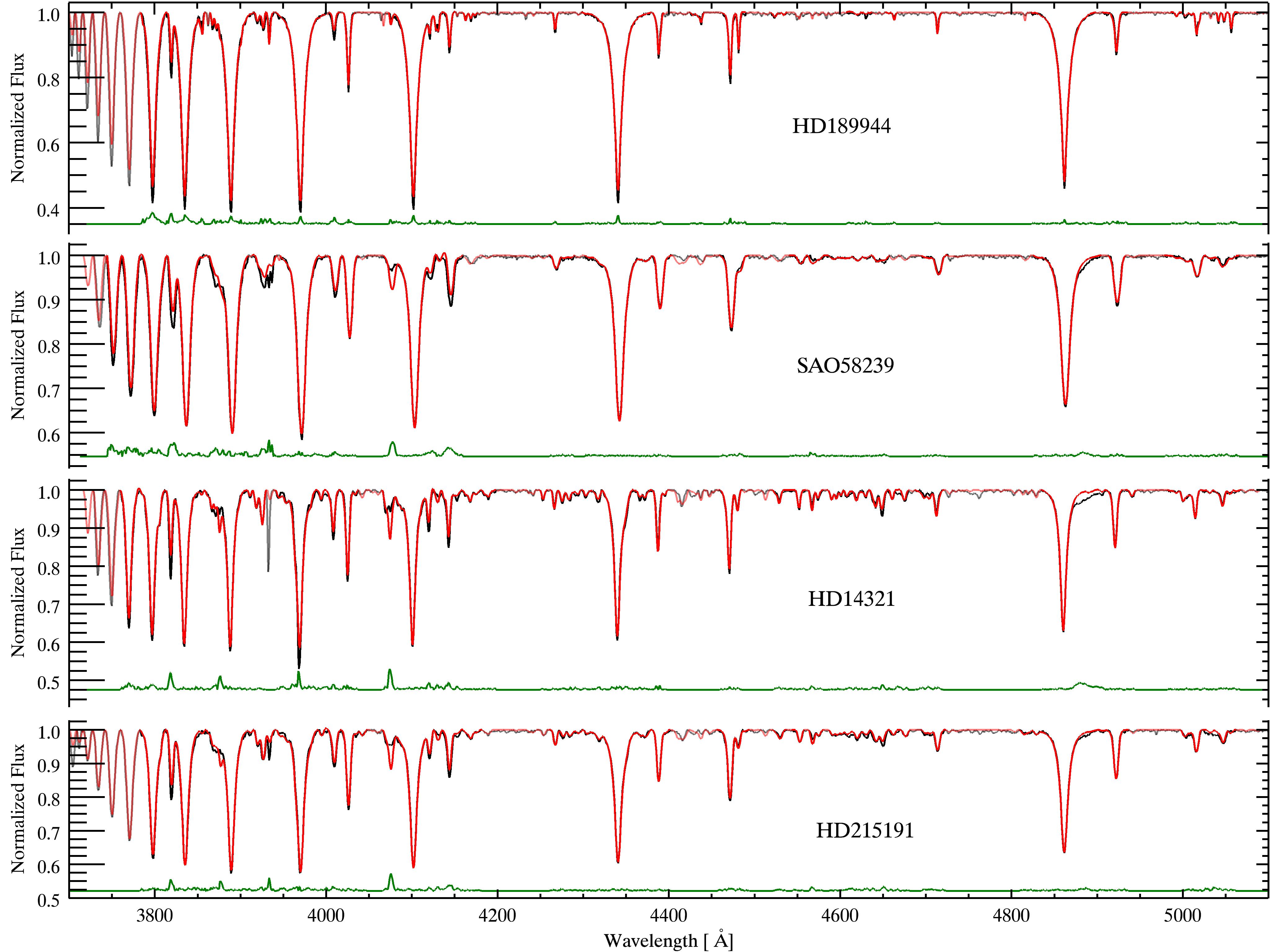}
  \caption{Examples of the best fits obtained for four B stars in our sample. These fits have the smallest averaged rms of our sample
(from top to bottom, the averaged rms are $0.0043$, $0.0042$, $0.0041$, and $0.0038$).
Observed spectra are in black, best synthetic spectra are in red, and the offset rms are in green. Regions of spectra not used for 
the analysis are in lighter colour. }
\label{fig:best3}
\end{figure*}

\begin{figure*}
\includegraphics[scale=0.12]{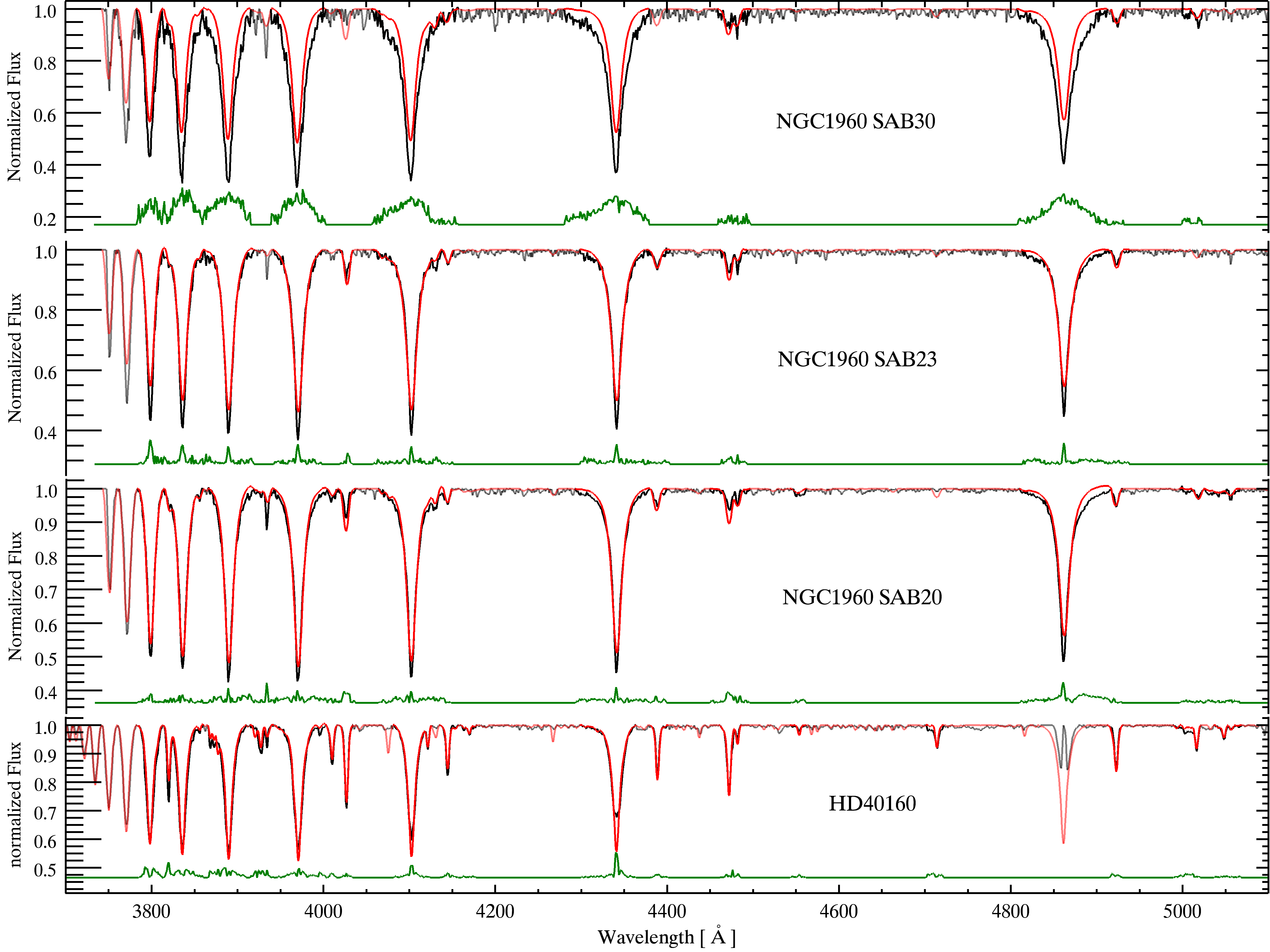}
  \caption{Examples of the worst fits obtained for four B stars in our sample. These fits have the highest averaged rms of our sample
(from top to bottom, the averaged rms are $0.05$, $0.011$, $0.0098$, and $0.0088$).
Colours are as in Figure \ref{fig:best3}. }
\label{fig:worse3}
\end{figure*}

We apply our method using most of the available lines for all 54 spectra. Figures~\ref{fig:best3} and~\ref{fig:worse3}
show 4 examples of the best fits obtained with a small and a high rms, respectively. All the other quality
fits are between these extreme examples. For all stars, the fit is globally good when considering the whole spectrum 
while it may vary from excellent to decent when considering each line individually. For instance, HD215191 in Figure~\ref{fig:best3}
has a global rms of $0.0038$, while individual lines have a rms between $0.0011$ and $0.0085$. 
This is because the method returns the best solution for all the lines at the same time \textit{and not the best solution for 
each line taken individually}. 

Concerning NGC1960 SAB30 (Fig. \ref{fig:worse3} upper panel), the poor quality of the fit (by far the worse case in our sample) is mainly 
due to the fact that this star has an effective temperature beyond the reach of our model. Its effective temperature is most 
likely around $13000$~K (HG06), while our grid can only go down to $15000$~K. This is also the case for NGC1960 SAB23 
(Fig. \ref{fig:worse3} second panel) but to a lesser extent, since its effective temperature seems to be closer to $15000$~K (HG06)
than for SAB30.

For HD40160 (Fig. \ref{fig:worse3} lower panel), we encountered some difficulty during the reduction process, due to the presence
of humidity spots on the CCD at the time of the observation, making the H$\beta$ line unusable, altering the shape of H$\gamma$ 
(i.e. a shallower centroid), and reducing the number of usable lines in general. We still performed the analysis of this star as an 
extreme test for the sensitivity of our method to the number and quality of the lines used. While our results for HD40160 are not 
extremely different from those found in the literature, they present one of the largest discrepancy of our results with other works, and 
therefore are not considered in the remaining of this paper. In particular, we find $T_{eff}=20010^{+290}_{-240}$~K,
$\log(g)=4.05^{+0.07}_{-0.05}$, $v\,\sin(i)=128^{+7}_{-8}$~km~s$^{-1}$, and $\xi=5^{+1}_{-2}$~km~s$^{-1}$, while HG06
found $T_{eff}=16453\pm250$~K, $\log(g)=3.735\pm0.027$, $v\,\sin(i)=115\pm35$~km~s$^{-1}$, and $\xi=2$~km~s$^{-1}$.

\begin{figure*}
\includegraphics[scale=0.12]{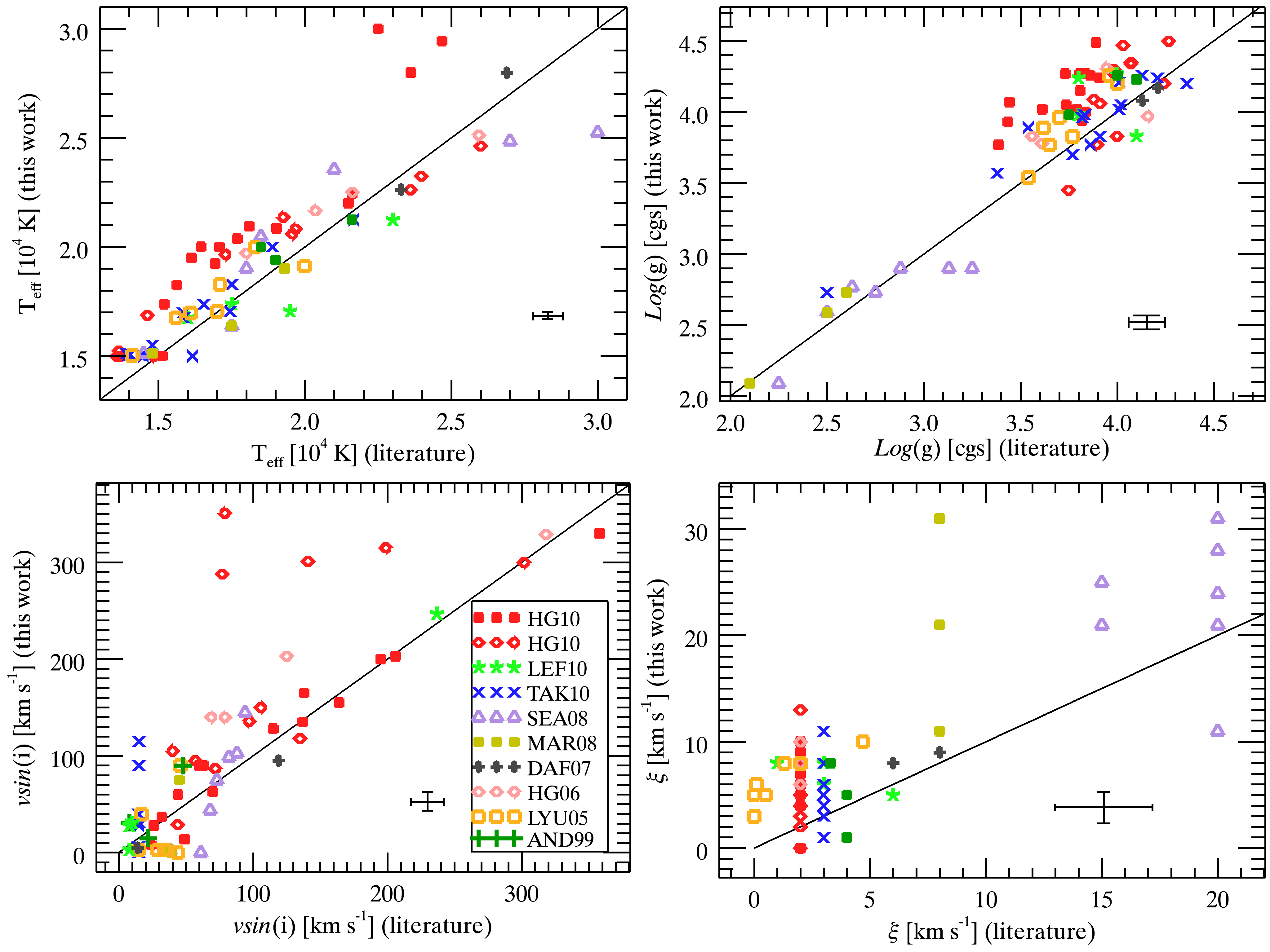}
  \caption{Comparison of our results for $T_{eff}$, $\log(g)$, $v\,\sin(i)$, and $\xi$ with other works from the literature.
  Atmospheric models from TLUSTY with the solar composition from Asplund et al (1999) have been used. 
Open diamonds are cluster stars and the other symbols are field stars. Mean error bars for these data are shown in black.
The black full line indicates a one-to-one match.
Note that some of our stars have been studied in more than one paper and thus appear multiple times in the graphics.
The abbreviations in the legend correspond to the following papers:
HG10 for \citet{Huang2010}; LEF10 \citet{Lefever2010}; TAK10 \citet{Takeda2010}; SEA08 \citet{Searle2008}; 
MAR08 \citet{Markova2008}; DAF07 \citet{Daflon2007}; HG06 \citet{Huang2006a,Huang2006b}; 
LYU05 \citet{Lyubimkov2005}; and AND99 for \citet{Andrievsky1999}.}
\label{fig:multiref}
\end{figure*}

\begin{figure*}
\includegraphics[scale=0.12]{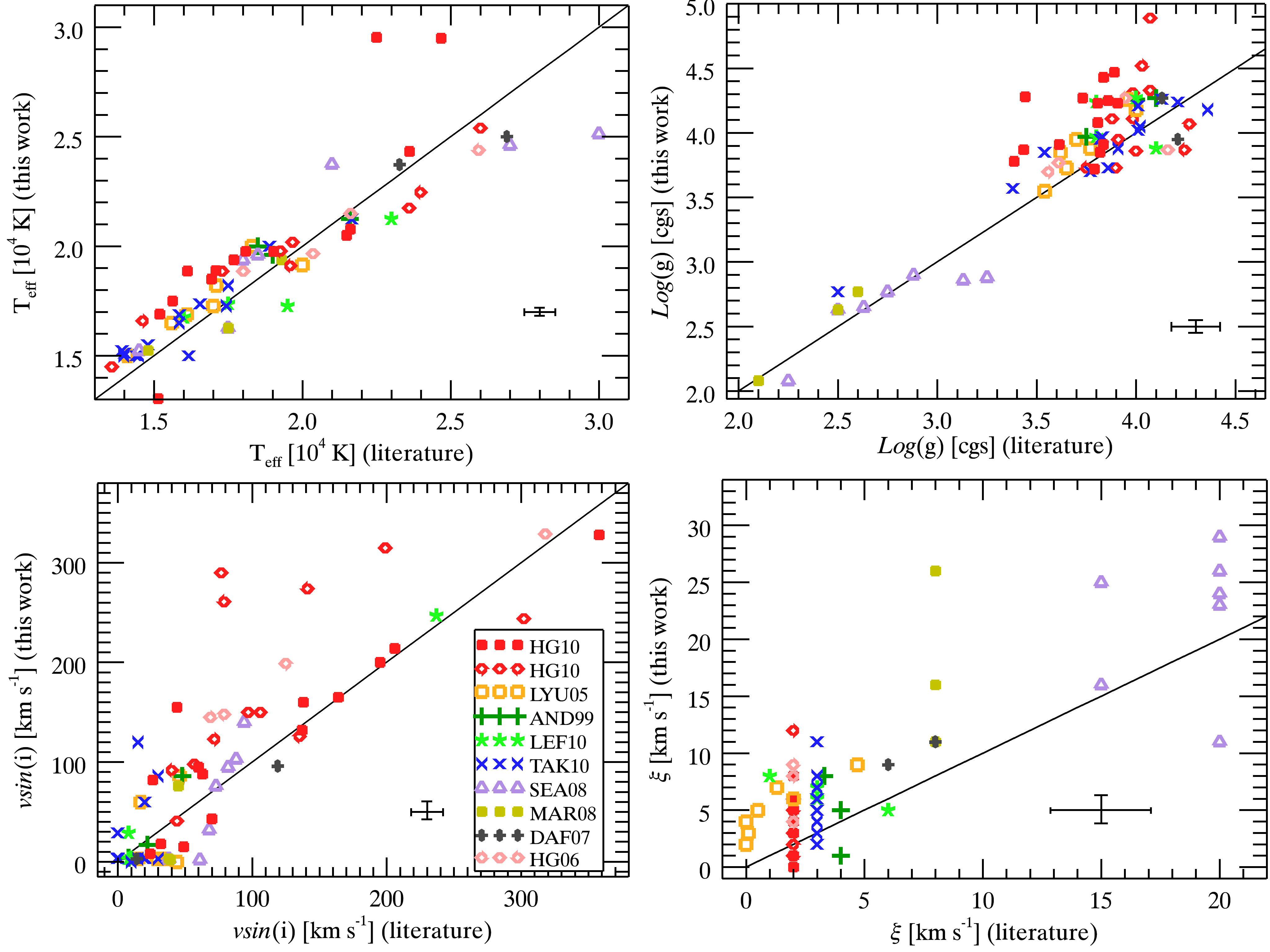}
  \caption{
  Comparison as in Figure \ref{fig:multiref} but with the individual chemical abundances fixed, in our analysis, 
  to the values given by other papers. The modified chemical elements are as follow: He and Si for a comparison 
  with LEF10; O and Ne for TAK10; C, N, and O for SEA08; He and Si for MAR08; C, N, O, Mg, Al, and Si for DAF07;
  He and Mg for LYU05; and C and N for a comparison with AND99. No individual chemical abundance modifications were considered for 
  the stars in common with HG10 and HG06, but we used the atmospheric model ATLAS9 with the solar abundances of \citet{Grevesse1998}
  in this case. Note that we find some stars with $T_{eff}< 15000$ K since the grid calculated with ATLAS9 has a broader
  range in $T_{eff}$ than our grid created with TLUSTY. Also, while HG10 derived $T_{eff}$ and $\log(g)$ using ATLAS9, these authors
  used TLUSTY to derive $v\,\sin(i)$, the comparison for this parameter is thus irrelevant here.}
\label{fig:indivref}
\end{figure*}

Figure \ref{fig:multiref} compares our results with those from other groups. All the numerical values can be found in Tables 
\ref{tab:pftab} (for field stars) and Table \ref{tab:pftabcl} (for cluster stars), where the uncertainties given by our method 
(as explained in $\S\,$\ref{param_estim}) are generally 
smaller than those found in the literature, because of the cumulative effect of our method (as explained in Fig. \ref{fig:methex}). 
While most of the time there is a good agreement between our results and those from other works, there are some important discrepancies. 
These are discussed in the following subsections.

\subsubsection{Chemical Abundances}

In this paper we wants to do a study of the impact of the spectral synthesis method used on the stellar parameters, 
therefore it is important that we first demonstrate that the discrepancies in Figure \ref{fig:multiref} are not associated to the chemical 
abundances considered by the different analysis. Some of the other works used here for comparison derived the abundances of a few chemical 
elements at the same time that they determined the basic stellar parameters, while we are considering a solar abundance, based 
on \citet{Asplund2009}, as the standard chemical composition for the models for all the stars in our sample. Nevertheless, the variations 
of the abundances found by the other authors are small, the overall metallicity is solar, and this distinction is not the main reason for 
the discrepancies found in Figure \ref{fig:multiref}. To demonstrate this, we repeat the analysis of the stars in our sample using the 
abundances derived by the other works. Specifically, we adopt the abundances of the following chemical elements: He and Si from LEF10 
(Table \ref{tab:lef_tab}); O and Ne from TAK10 (Table \ref{tab:tak_tab}); C, N and O from SEA08 (Table \ref{tab:Sear_tab}); He and Si 
from MAR08 (Table \ref{tab:mark_tab}); C, N, O, Mg, Al, and Si for DAF07 (Table \ref{tab:daf_tab}); He and Mg from LYU05 
(Table \ref{tab:Lyu_tab}); and C and N from AND99 (Table \ref{tab:and_tab}), for the stars that we have in common. HG10 and HG06
did not derive the abundances for any specific elements but they used the atmospheric model ATLAS9 with the solar abundances from 
\citet{Grevesse1998}, as we also redo for the corresponding stars (Tables \ref{tab:pftabBK} and \ref{tab:pftabclBK}). 
Note that HG10 and HG06 derived $T_{eff}$ and $\log(g)$ using ATLAS9, but determined $v\,\sin(i)$ using TLUSTY models and 
as such, one should consider only Figure \ref{fig:multiref} for a proper comparison of $v\,\sin(i)$.
Figure \ref{fig:indivref} shows the comparison of our new results, with the other papers, obtained when we are using their abundance data. 
When comparing Figures~\ref{fig:multiref} and~\ref{fig:indivref}, we see a small and marginal difference between the two set of results 
(except for the stars studied by HG10 which we discuss in more details in $\S\,$\ref{restefflog}. We believe that the reason for the small 
change between the two figures is mainly related to the fact that the element abundances derived by the other authors were tailored to a 
few lines only and have little impact on the other parameters obtained simultaneously by fitting the whole spectrum as is done with 
our method. As discussed in the following subsections, we find that the impact of the method used and the choice of the diagnostic lines 
is more important to explain the discrepancies seen in Figures \ref{fig:multiref} or~\ref{fig:indivref}. A complete analysis with our method, 
including abundances determination and their effect on the basic stellar parameters, will be done in a following paper.

\subsubsection{Effective Temperature and Surface Gravity}\label{restefflog}

\begin{figure*}
\includegraphics[scale=0.1]{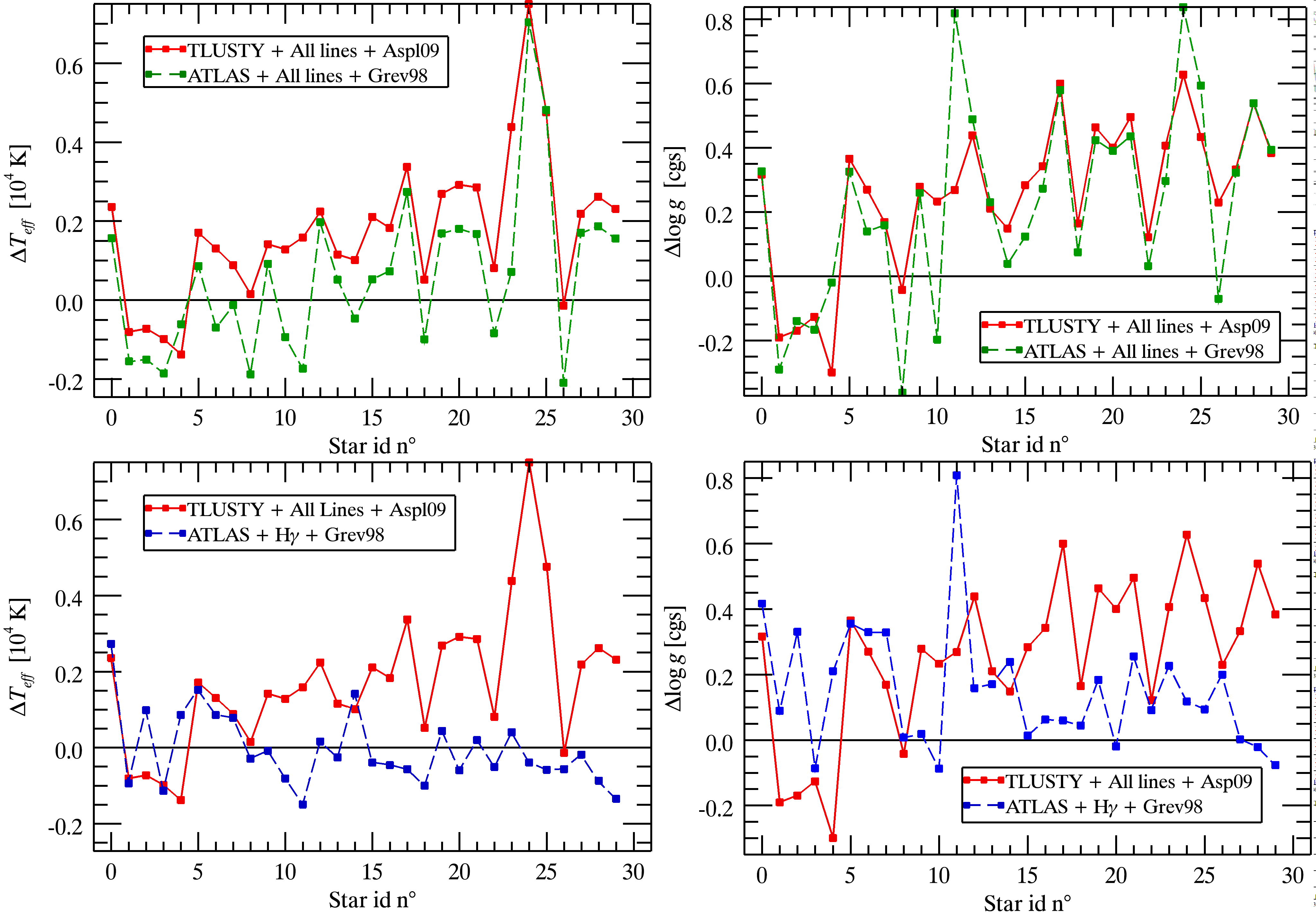}
  \caption{$T_{eff}$ and $\log(g)$ variations for different models and different number of lines for the 30 stars in common 
  with the sample of \citet{Huang2010} and \citet{Huang2006a}. The variations are defined as 
  $\Delta T_{eff}= T_{eff} (this\,work) - T_{eff} (Huang~et~al.)$ and $\Delta \log(g)= \log(g) (this\,work) - \log(g) (Huang~et~al.)$.
  \textit{Upper panels}: Variations when considering all the lines but different atmospheric models and solar
    chemical composition. \textit{Lower panels}: Variations when considering all the available lines or only H$\gamma$.}
\label{fig:Delta_teff_logg}
\end{figure*}

Figure \ref{fig:multiref}, or \ref{fig:indivref}, often shows a large scattering for the effective temperature and the surface gravity. 
Among others, we see a quasi-systematic difference between our estimates of $T_{eff}$ and $\log(g)$ and those of HG10 and HG06. 
We believe that the main reason for this is related to the fact that these authors used only H$\gamma$ as a diagnostic line. 
For the 30 stars that we have in common with HG10 and HG06, we perform our analysis using the
atmospheric model ATLAS9 and the solar chemical composition of \citet{Grevesse1998} with
all the available spectral lines (these results are presented in Fig. \ref{fig:indivref}), but also, with only H$\gamma$ as a 
diagnostic line. Figure \ref{fig:Delta_teff_logg} shows for the 30 stars
the effect on $T_{eff}$ and $\log(g)$ of the number of diagnostic lines used, as well as the effect of the different models. 
From the upper panels of Figure \ref{fig:Delta_teff_logg}, we see that a change of atmospheric
model or chemical composition (TLUSTY with the solar composition of \citealt{Asplund2009} in red versus ATLAS9 and the solar 
composition of \citealt{Grevesse1998} in green) has little impact on the
discrepancies between our results and those of HG10 and HG06: the discrepancies 
remain important. Whereas the lower panels shows that considering only H$\gamma$ as a diagnostic
line drastically reduces the variations in $T_{eff}$, and greatly affect those in $\log(g)$. For instance, the 
3 stars in Figure \ref{fig:multiref} with a large difference in $T_{eff}$, when compared to HG10, are the most extreme
examples of this effect (they are the stars number 23, 24, and 25 in Figure \ref{fig:Delta_teff_logg}).
It is clear in this case that the number and choice of diagnostic lines used has a greater impact on the determination 
of $T_{eff}$ and $\log(g)$ than the choice of the atmospheric model or the solar chemical composition.

\begin{figure}
\includegraphics[scale=0.06]{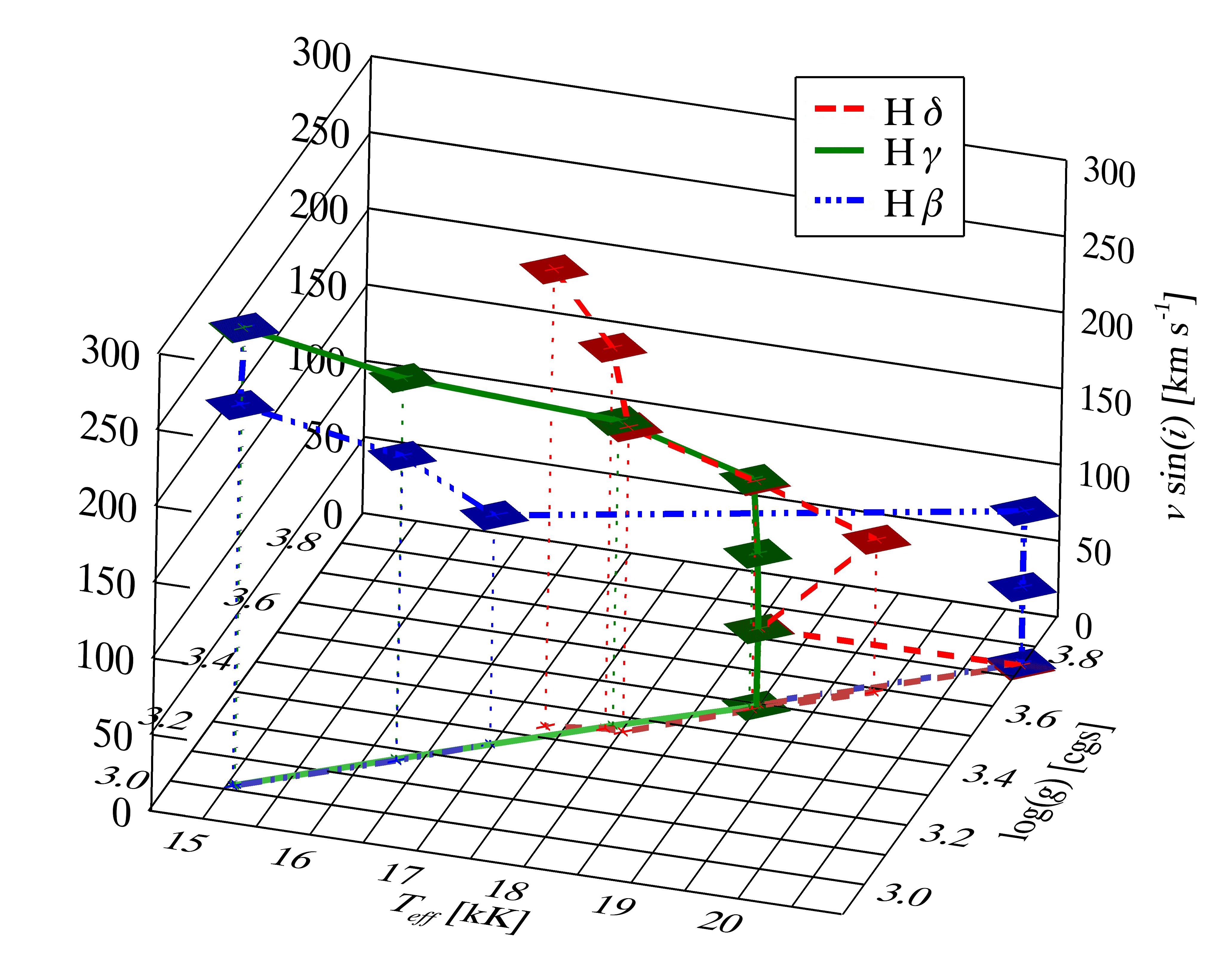}
  \caption{The most probable pair [$T_{eff}$, $\log(g )$] as a function of $v\,\sin(i)$
for different lines (as indicated in the plot) for the artificial spectrum. The parameters used to create the artificial 
spectrum are: $T_{eff}=21350$~K, $\log(g)=3.8$, $\xi=3$~km~s$^{-1}$, and $v\,\sin(i)=15$~km~s$^{-1}$.
The behaviour as a function of $v\,\sin(i)$ is shown projected in the plane [$T_{eff}$, $\log(g)$]. 
The likelihoods are marginalized over all $\xi$ values. 
Note in particular the most probable pair [$T_{eff}19000$~K, $\log(g)=3.5$] given by H$\gamma$ when $v\,\sin(i)\leq150$~km~s$^{-1}$.
} 
\label{fig:HbetHgam}
\end{figure}

Furthermore, the choice of the diagnostic line, when only one is used, also affects the results. 
Figure \ref{fig:HbetHgam} shows the best values of $T_{eff}$ and $\log(g)$ that we obtain for different values of $v\,\sin(i)$ 
when only H$\beta$, H$\gamma$, or H$\delta$ is considered for the analysis. This analysis is done using an artificial spectra 
with $T_{eff}=21350$~K, $\log(g)=3.8$, $\xi=3$~km~s$^{-1}$, and $v\,\sin(i)=15$~km~s$^{-1}$. In Figure \ref{fig:HbetHgam}, we can 
see that each line has a different sensitivity over $v\,\sin(i)$ and that two lines rarely give the same results for a given 
$v\,\sin(i)$. Moreover, $T_{eff}$ and $\log(g)$ given by H$\gamma$ are lower than the real values by 2350 K and 0.3 dex, respectively,
when $v\,\sin(i)$ is close to its real value. These differences are of the order of the quasi-systematic difference seen between 
HG10, HG06, and our work.

When comparing our results for the evolved stars with those obtained by \citet{Searle2008} in Figure \ref{fig:multiref}, 
we find similar temperatures for the stars below 20000~K, and less scatter than \citet{Searle2008} for the stars above 
(the effective temperature goes from 23500 K to roughly 25000 K in our work, while the authors found temperatures
going from 21000 K to 30000 K). The $\log(g)$ values are, within the uncertainties, quite identical, with $\log(g)$ below 3 in our case.
We also find higher microturbulence velocities and a larger scatter for the rotational velocity. 
Note that performing our analysis with the C, N, and O abundances derived by \citet{Searle2008} has little to
no impact on the observed discrepancies (as shown when comparing Figures \ref{fig:multiref} and \ref{fig:indivref}). 
The reason for the differences between our results and those of \citet{Searle2008} may be related to two important effects:

\noindent 1) First of all, \citet{Searle2008} used a combination of TLUSTY and CMFGEN in order to include wind effects which we are 
not doing. For instance, in B supergiant, the Balmer lines could suffer from wind contamination, where photons emitted by hydrogen 
atoms in the wind ``fill in'' the absorption profiles thus making them shallower than any profiles calculated with a photospheric 
model \citep{Searle2008}. It is especially true for H$\alpha$ and H$\beta$ and to a lesser extent for H$\gamma$ and the other Balmer 
lines. Other photospheric lines can also be influenced by winds, such as the Si\,{\tt IV}\,$\lambda$4089 line and the 
Si\,{\tt III}\,$\lambda\lambda$4552,4568,4575 multiplet \citep{2003ApJ...588.1039H,Dufton2005,Searle2008}. But since this influence 
seems to be model dependant, it is therefore difficult to correctly quantify it. Nevertheless, it is not really surprising that
among the 8 program stars that we have in common with \citet{Searle2008}, the 3 stars that exhibit the strongest differences in
$T_{eff}$ and $\log(g)$ are the ones with the strongest mass-loss rate (according to \citealt{Searle2008} measurements). 
These 3 stars are HD192660, HD213087, and HD190066. Respectively, their mass-loss rates are $5$, $0.7$, and 
$0.7\times10^{-6}M_{\odot}$~$yr^{-1}$, their effective temperature differences are $4290$, $3250$, and $2370$ K, 
and their $\log(g)$ differences are $0.3$, $0.27$, and $0.07$ dex.

\begin{figure*}
\includegraphics[scale=0.12]{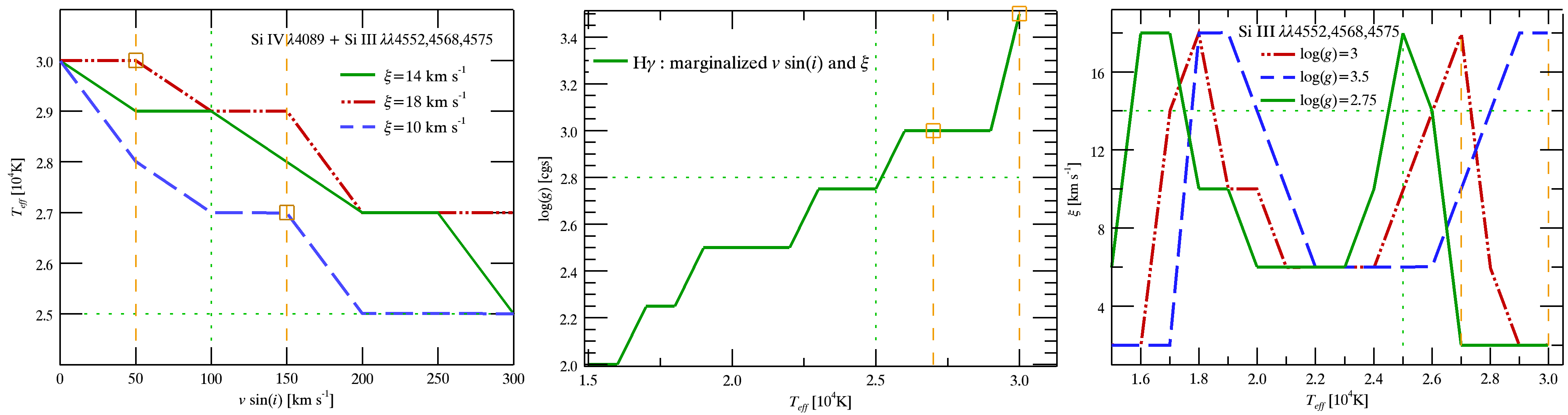}
  \caption{Illustration of the interdependency between $T_{eff}$, $\log(g)$, and $\xi$ when individual diagnostic lines are considered
  in an iterative method. The parameters of the artificial spectrum used in this exemple are:  $T_{eff}=25000$~K, $\log(g)=2.8$,
  $v\,\sin(i)=100$~km~s$^{-1}$, and $\xi=14$~km~s$^{-1}$. \textit{Left Panel}: Most probable effective temperature as a function of
  $v\,\sin(i)$ for 3 different microturbulence velocities using Si\,{\tt IV}\,$\lambda$4089 and Si\,{\tt III}\,$\lambda\lambda$4552,4568,4575.
  Orange squares and thin orange dashed lines indicate the extreme solutions found when underestimating and overestimating $v\,\sin(i)$ 
  compared to the parameters used to create the artificial spectrum (thin green dotted lines in all panels). 
  \textit{Middle panel}: Most probable $\log(g)$ values as a function of $T_{eff}$, after marginalizing over $v\,\sin(i)$
  and $\xi$ while using H$\gamma$. Orange squares and orange dashed lines indicate the solution found when overestimating and underestimating
  $T_{eff}$ from the left panel. 
  \textit{Right Panel}: Most probable microturbulence velocities as a function of $T_{eff}$, after marginalizing over 
  $v\,\sin(i)$, for 3 different $\log(g)$ estimates. The full green, dashed blue, and dot-dashed red lines represent the possible values of 
  $\log(g)$ found in the middle panel when overestimating and underestimating
  $T_{eff}$. The thin orange dashed lines indicate overestimated and underestimated $T_{eff}$ from the left panel.}
\label{fig:Itertlvt}
\end{figure*}

\noindent 2) Furthermore, \citet{Searle2008} used, in an iterative way, different diagnostic lines for different parameters each time, while fixing the other 
parameters and adopting a rotational velocity found in the literature. While vastly used, due to its simplicity and swiftness, this iterative 
method may provide a local maximum in the probability space rather than a global maximum because it considers each parameter independently. 
For instance, \citet{Searle2008} used Si\,{\tt IV}\,$\lambda$4089 and Si\,{\tt III}\,$\lambda\lambda$4552,4568,4575 as primary diagnostic 
lines for the effective temperature with fixed values of $v\,\sin(i)$ and $\xi$. Then they used the effective temperature obtained and the 
H$\gamma$ line to determine $\log(g)$. And finally they adopted their pair [$T_{eff}$, $\log(g)$], and a value of $v\,\sin(i)$ found in the 
literature, to determine a new microturbulence velocity using the Si\,{\tt III}\,$\lambda\lambda$4552,4568,4575 lines. The whole process 
was repeated until small changes in the parameters were seen. With this iterative method, the effective temperature returned by the silicon 
lines greatly depends on the values used for the microturbulence and rotational velocities. And $\log(g)$ obtained from H$\gamma$ also 
depends on the effective temperature. Finally, the microturbulence is significantly affected by $T_{eff}$ and $\log(g)$. 
To illustrate the impact of an iterative method, we create another artificial stellar spectrum with the following 
specifications: $T_{eff}=25000$~K, $\log(g)=2.8$, $v\,\sin(i)=100$~km~s$^{-1}$, and $\xi=14$~km~s$^{-1}$. We then calculate the posterior 
probability for the 4 silicon lines from this artificial spectrum with the \textit{BSTART} grid. After the marginalization over the $\log(g)$ 
domain, we find the most probable effective temperature for each given value of $\xi$ and $v\,\sin(i)$ (Fig. \ref{fig:Itertlvt}, left panel). 
We then perform the same operation for H$\gamma$ and marginalize over $v\,\sin(i)$ and $\xi$ (since both parameters have little influence
on the returned $\log(g)$ value, showing the marginalized probability is enough in this case) to obtain the most probable $\log(g)$ 
estimate given each value of $T_{eff}$ (Fig. \ref{fig:Itertlvt}, middle panel). 
And finally we redo the same analysis, but this time for only the 3 lines Si\,{\tt III}\,$\lambda\lambda$4552,4568,4575 and with a 
marginalization over $v\,\sin(i)$ in order to find the most probable value of $\xi$ for each value of $T_{eff}$ and $\log(g)$ 
(Fig. \ref{fig:Itertlvt}, right panel). To conclude, we can see that the chosen silicon lines almost always overestimate the effective 
temperature (except when $\xi=10$~km~s$^{-1}$
and $v\,\sin(i)\geq200$~km~s$^{-1}$). Figure~\ref{fig:Itertlvt} also shows a clear anti-correlation: the effective temperature decreases 
as the rotational velocity increases. Thus if we were to underestimate (respectively overestimate) $v\,\sin(i)$, we would overestimate 
(respectively underestimate) $T_{eff}$. As $\log(g)$ increases with the effective temperature, we would therefore overestimate (respectively underestimate) 
$\log(g)$. Finally, we would overestimate or underestimate the microturbulence velocity depending on how we estimate $\log(g)$.
This behaviour is what we observe between our results and those of \citet{Searle2008} for the stars hotter than $18000$~K. 
For the cooler stars, \citet{Searle2008} used Si\,{\tt III}\,$\lambda\lambda$4128,4130 rather than Si\,{\tt IV}\,$\lambda$4089, which seems
to induce a smaller effective temperature discrepancy with our method.

Results from the other authors in our comparison list are in a reasonably good agreement with our results even though a wide variety 
of technics and methods were used.
\citet{Takeda2010}, \citet{Lyubimkov2005}, and \citet{Andrievsky1999} used the \textit{uvby$\gamma$} photometric system to derive $T_{eff}$ 
and $\log(g)$. Our results are quite consistent with theirs. \citet{Daflon2007} derived $T_{eff}$ using the \textit{UVB} photometric system 
and obtained $\log(g)$ by adjusting the observed H$\gamma$ line wings with models from ATLAS9. Note that we do not observe the same 
discrepancies in $\log(g)$ with \citet{Daflon2007}, as seen between our results and those of HG10, since on the one hand, 
\citet{Daflon2007} used $T_{eff}$ values close to ours, and on the other hand, they did not fit the core of the H$\gamma$ line
where non-LTE effects are stronger. \citet{Markova2008} used a similar approach than \citet{Searle2008}: models including stellar wind, 
$\log(g)$ from Balmer lines, and $T_{eff}$ 
from Si lines. Our results agree well with theirs since the stars we have in common have weak stellar winds and $T_{eff}$ below or around
18000 K. Finally, \citet{Lefever2010} used a semi-automated iterative method which is close to our simultaneous approach because groups of
parameters are derived using a large number of lines simultaneously. Our respective results are in a reasonable agreement considering that
\citet{Lefever2010} used models including stellar wind (FASTWIND) and an iterative method.

\subsubsection{Projected Rotational Velocity}

 \begin{figure}
\includegraphics[scale=0.08]{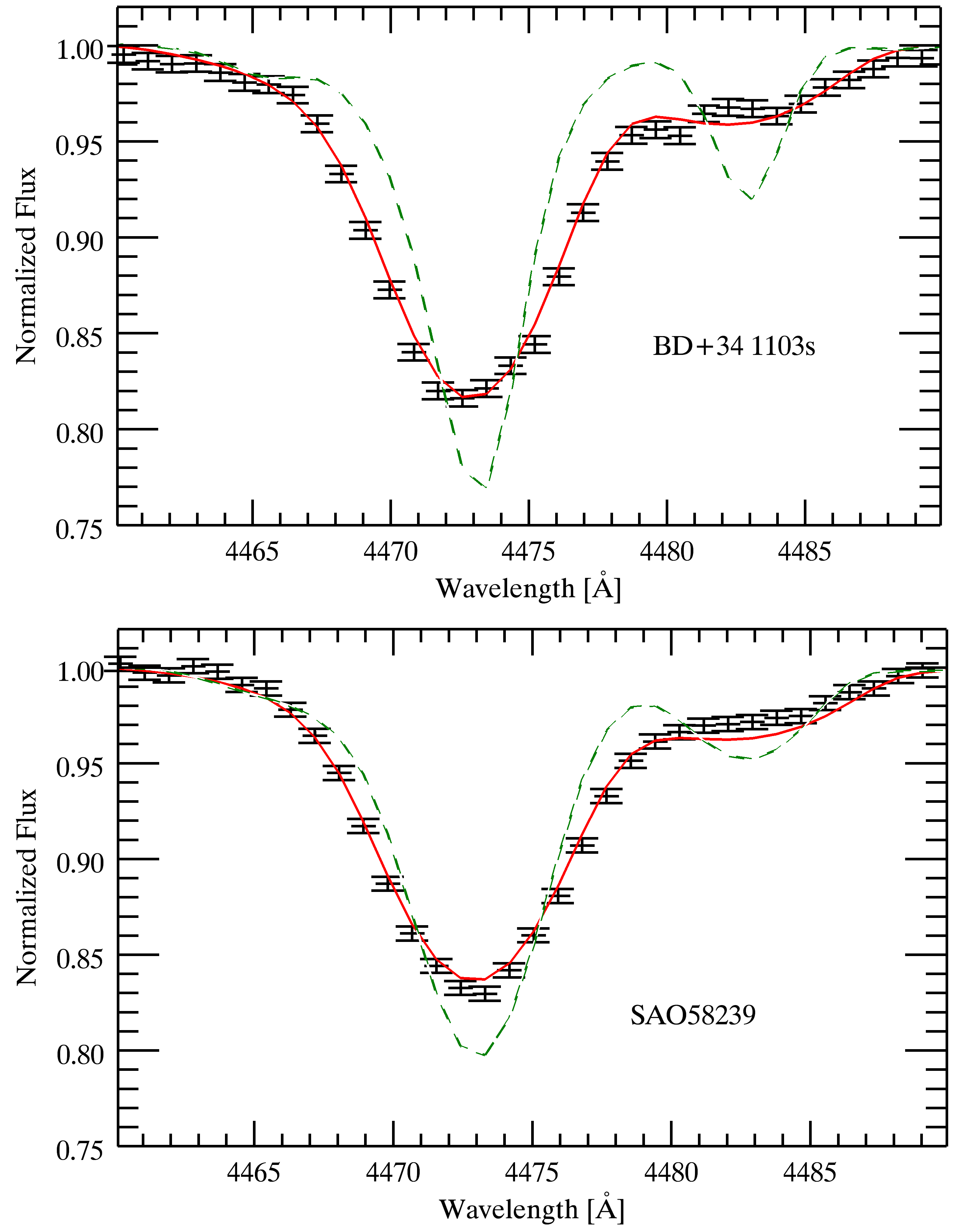}
  \caption{Comparison of the best fits for the lines He\,{\tt I}\,$\lambda$4471 and Mg\,{\tt II}\,$\lambda$4481  
  for 2 stars with large rotational velocities. Individual data points with error bars represent our observations.
  In full red: models based on our best set of parameters. In dashed green: models using the parameters of HG10. 
  The $v\,\sin(i)$ values are: 288~km~s$^{-1}$ (red) and 77~km~s$^{-1}$ (green) for BD+34$\,$1103s; and 
  315~km~s$^{-1}$ (red) and 199~km~s$^-1$ (green) for SAO58329.
  The respective complete sets of parameters are listed in Table \ref{tab:pftabcl}.
  }
\label{fig:highvr}
\end{figure}

A substantial difference in Figure \ref{fig:multiref},  between our results and those published in the literature, 
concerns $v\,\sin(i)$. The most important discrepancy is found for 4 cluster stars that are also in the sample of HG10 
(as mentioned earlier, they also used TLUSTY to derive $v\,\sin(i)$). We indeed 
derive higher $v\,\sin(i)$ values (around 300~km~s$^{-1}$) by a factor of at least $1.5$ for BD+34$\,$1103s, NGC1960 SAB20, 
NGC1960 SAB30, and SAO58239. Of these stars, SAB20 and SAB30 have effective temperatures below our limit of 
15000 K and the lowest SNR (around 200  and 75 respectively) objects of our sample. Consequently, the best fits for these 
stars are among the worst in our sample (Fig. \ref{fig:worse3}) even when using ATLAS9 models, we thus consider that the associated results 
are not reliable. On the contrary, the other two stars, BD+34$\,$1103s and SAO58239, have excellent SNR and 
 rms (see Fig. \ref{fig:best3} for SAO58239). Here, the differences in $v\,\sin(i)$ cannot be explained simply either in terms of the 
 number
  \begin{figure*}
\includegraphics[scale=0.1]{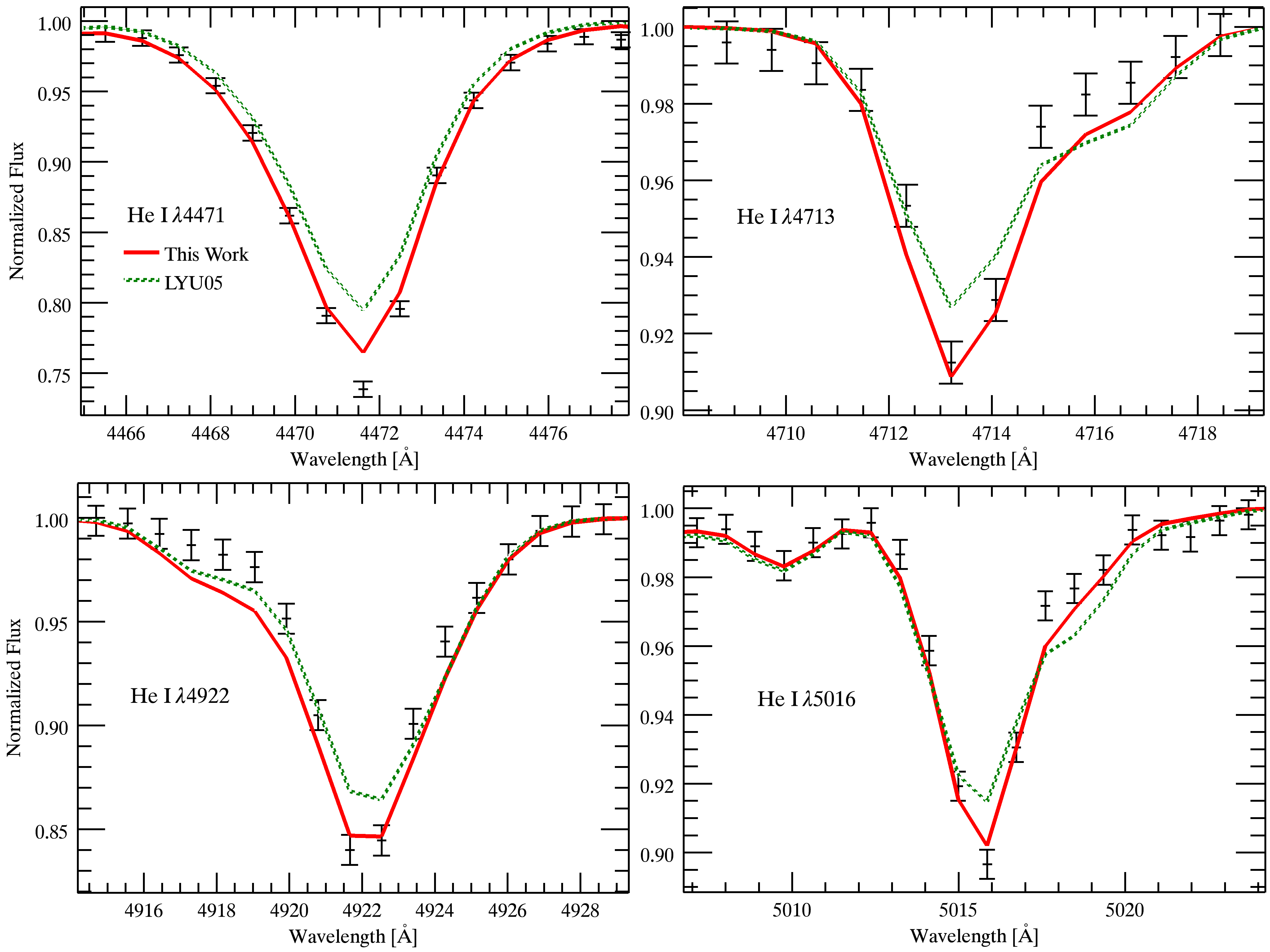}
  \caption{Comparison of the fits for 4 helium lines obtained with 
  our best set of parameters and those from \citet{Lyubimkov2005}, for HD184171.
  Individual data points with error bars represent our observations. In full red: our best model with $T_{eff}=16890$~K, $\log(g)=3.89$,
  $\xi=3$~km~s$^{-1}$, and $v\,\sin(i)=3$~km~s$^{-1}$. In dotted green: model using the parameters of \citet{Lyubimkov2005}: $T_{eff}=16100$~K, 
  $\log(g)=3.62$, $\xi=0$~km~s$^{-1}$, and $v\,\sin(i)=29$~km~s$^{-1}$.
  }
\label{fig:HD184171}
\end{figure*}
 or the choice of lines, or in terms of the atmospheric models used, since HG10 also used TLUSTY and SYNSPEC to determine $v\,\sin(i)$ 
 and we could not reproduce their results using the same lines. One plausible explanation would be the difference in the treatment 
 of $v\,\sin(i)$ (HG10 used a grid of models based on a three-dimensional parameterization of $T_{eff}$, $\log(g)$, and the cosine of 
 the angle between the surface normal and the line of sight) but, in this case, there should be a systematic effect between our derived 
 values and theirs for all the cluster stars that we have in common, which we do not observe. Moreover, when we reproduce the lines 
 He\,{\tt I}\,$\lambda$4481 and Mg\,{\tt II}\,$\lambda$4481, used by HG10 to derive $v\,\sin(i)$, we clearly obtain an excellent fit 
 of our own data with our parameters (Fig. \ref{fig:highvr}). We are thus confident of our derived parameters for these two stars even though
 we have no clear explaination for these discrepancies.
 
Interestingly, while the differences between our results for $v\,\sin(i)$ and those of HG10 and HG06 are generally greater
for cluster stars than for field stars, and even though our sample is not large, we still find the same conclusion
as HG06, i.e. cluster stars are globally faster rotators than field stars. From our results, cluster stars are
even faster rotators.

Another issue concerning $v\,\sin(i)$ in Figure \ref{fig:multiref}, or \ref{fig:indivref}, appears at relatively low rotational velocities 
($v\,\sin(i)\leqslant100$~km~s$^{-1}$). In this case, there is a rather large scatter between our results and those from other authors. 
Some of the largest differences can either be explained by our limited range in temperature (for example, \citealt{Takeda2010} 
found $v\,\sin(i)=15$~km~s$^{-1}$ for HD185330, while we 
find $v\,\sin(i)=115$~km~s$^{-1}$ because the effective temperature is most certainly below 15000~K), or by the parameters and spectral
lines considered. For instance, the $v\,\sin(i)$ values of \citet{Lyubimkov2005} are mostly between 15 and 50~km~s$^{-1}$, while ours 
are between 0 and 10~km~s$^{-1}$ with an uncertainty around 10~km~s$^{-1}$ for the same stars. \citet{Lyubimkov2005} derived their 
velocities using fixed values of $T_{eff}$ and $\log (g)$, free values of $v\,\sin(i)$, $\xi$, and helium abundances, and considered 6 
helium lines, while we use solar helium abundance, free values of $T_{eff}$, $\log (g)$, $v\,\sin(i)$, and $\xi$, and more spectral 
lines. While helium abundance has an obvious impact on the shape of helium lines, relatively small changes in  $T_{eff}$, $\log (g)$,
and $\xi$ also have an influence on the resulting $v\,\sin(i)$ value. For instance, when considering HD184171 with $T_{eff}=16100$~K 
and $\log g=3.62$, \citet{Lyubimkov2005} found a solar helium abundance, $v\,\sin(i)=29$~km~s$^{-1}$, and $\xi=0$~km~s$^{-1}$, 
while we obtain $T_{eff}=16890$~K, $\log g=3.89$, $v\,\sin(i)=3$~km~s$^{-1}$, and $\xi=3$~km~s$^{-1}$ for a solar helium abundance. 
Figure \ref{fig:HD184171} shows the models for these 2 sets of parameters for 4 helium lines (that are among the 6 helium lines 
used by \citealt{Lyubimkov2005}). From this figure, it is clear that, with our set of parameters, a higher value of $v\,\sin(i)$ would 
not improve the fits. Note also that our parameters are derived from the analysis of more than these 4 helium lines (27 in this case), and as 
such are more constrained, thus explaining the small uncertainties. Indeed, if we perform an analysis using only these 4 helium lines, 
with $T_{eff}$, $\log (g)$, and $\xi$ fixed to their best values, the uncertainty for $v\,\sin(i)$ goes roughly from $\pm$10 to $\pm$20 
thus reducing the discrepancy between our results and those of \citet{Lyubimkov2005}.

\begin{figure}
\includegraphics[scale=0.07]{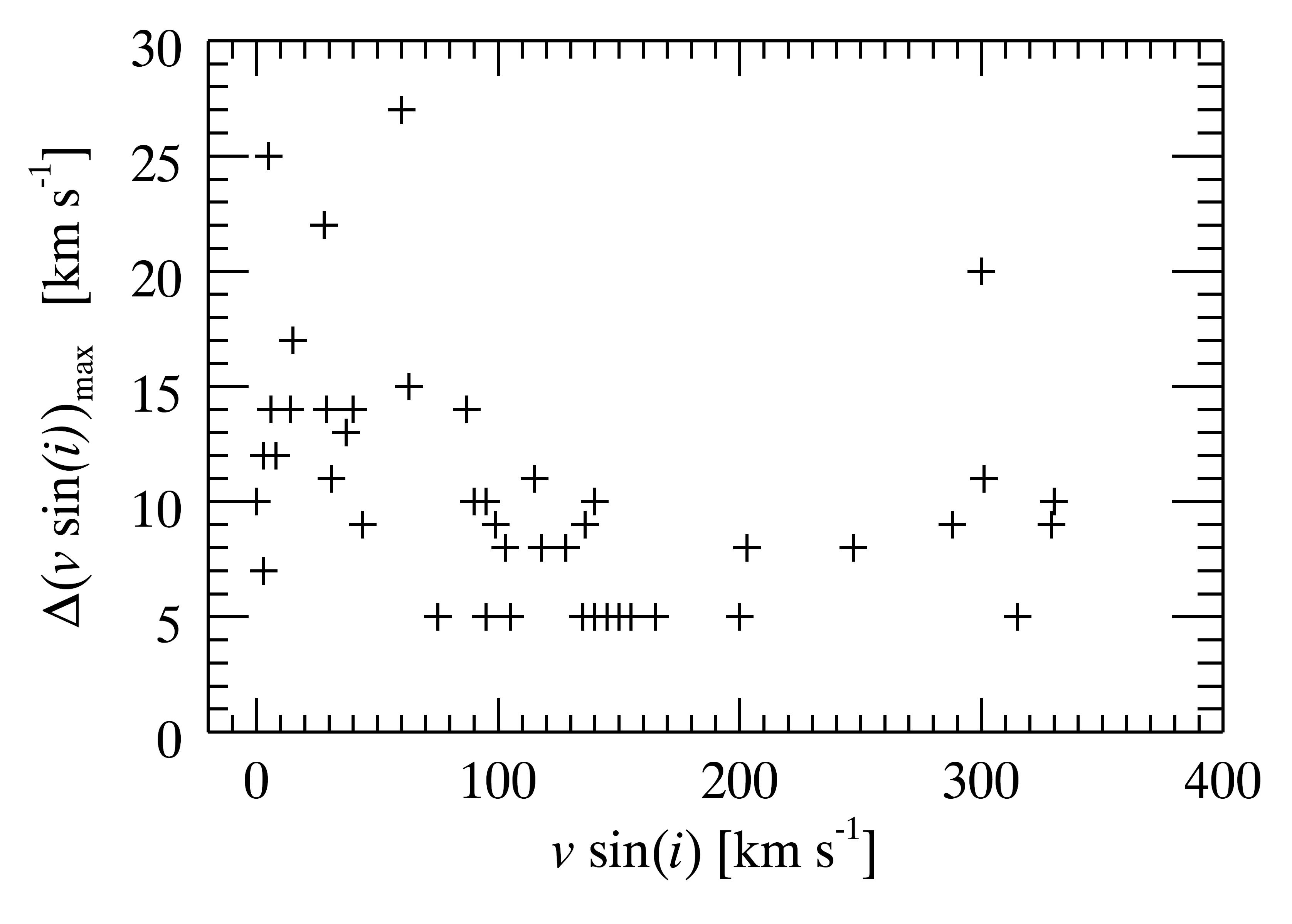}
  \caption{Maximum uncertainty for $v\,\sin(i)$ as a function of $v\,\sin(i)$ for our whole sample.}
\label{fig:dvsini}
\end{figure}

There is also an interesting point concerning low rotational velocities and their respective uncertainties. Indeed, HG10
noted that depending on the resolution of a given spectrum, and for $v\,\sin(i)$ less than half the FWHM of 
 the instrumental broadening function, the instrumental broadening dominates the rotational broadening, implying 
uncertainties of the order of $\pm$FWHM$/2$ for low $v\,\sin(i)$ values. In the work of HG10, FWHM $\approx0.8$ and $1.5$~\r{A} resulting 
in uncertainties of 29 and 48~km~s$^{-1}$, respectively, for $v\,\sin(i)$ below these values. Our data have a
FWHM of $2.3$~\r{A}, and thus the theoretical limit is around 80~km~s$^{-1}$. But, as our method cumulates the constraints of 
a large number of lines, it allows us to significantly reduce this limit. Figure \ref{fig:dvsini} shows, for each star, the maximum
uncertainty of $v\,\sin(i)$ as a function of $v\,\sin(i)$. Here, we see that the method generally and naturally returns larger
uncertainties for $v\,\sin(i)\leq 80$~km~s$^{-1}$ than for higher values, but with a maximum uncertainty (for low $v\,\sin(i)$) of
roughly 30~km~s$^{-1}$ well below the limit of 80~km~s$^{-1}$. As a further demonstration, we apply our method on higher 
spectral resolution data of two stars in our sample, obtained with a different grating at the Observatoire du Mont-M\'egantic (in February 2011).
These two stars are HD77770 and HD44700, whose spectra have an effective resolution of $0.9$~\r{A} and a wavelength range spanning from $3600$ 
to $4400$~\r{A}. The results for $T_{eff}$, $\log (g)$, and $\xi$ obtained for these two stars are the same (within the uncertainties)
at moderate and high resolution. But for $v\,\sin(i)$, HD77770 gives $0_{-0}^{+13}$~km~s$^{-1}$ at high resolution, 
rather than $14_{-14}^{+6}$~km~s$^{-1}$ at moderate resolution, and HD44700 gives $v\,\sin(i)=4_{-4}^{+10}$~km~s$^{-1}$, 
at high resolution, rather than $28_{-22}^{+18}$~km~s$^{-1}$ at moderate resolution.\\ 

\subsubsection{Microturbulence Velocity}
  
Finally, one of the most obvious difference in Figure \ref{fig:multiref}, or \ref{fig:indivref}, lies with the microturbulence velocity but 
it is simply explained by the fact that, in most works, the microturbulence velocity is not a free parameter. Usually $\xi$ is fixed at 
about $3$~km~s$^{-1}$ and $10$~km~s$^{-1}$ for main-sequence stars and for evolved stars, respectively \citep[HG10;][]{Takeda2010}. 
And, when the microturbulence velocity is not fixed, it is estimated with only one or two lines in the spectrum \citep{Lyubimkov2005} 
or with only one species \citep{Lefever2010}. When considering all the available lines to fit the microturbulence velocity simultaneously 
with the other parameters, we find a higher estimate than what is generally used. 
We obtain a mean microturbulence velocity of 
$6.1$~km~s$^{-1}$ with a standard deviation of $2.6$~km~s$^{-1}$ for the main-sequence stars, and a mean value of $21.4$~km~s$^{-1}$ 
with a standard deviation of $7.6$~km~s$^{-1}$ for the evolved stars. Note that our results are given for solar abundances, but even 
though we find slightly lower values of microturbulence velocity (by nearly 2~km~s$^{-1}$) when we use the abundances derived by
the various authors, our values are still higher than those from the literature.  

\begin{figure}
\includegraphics[scale=0.06]{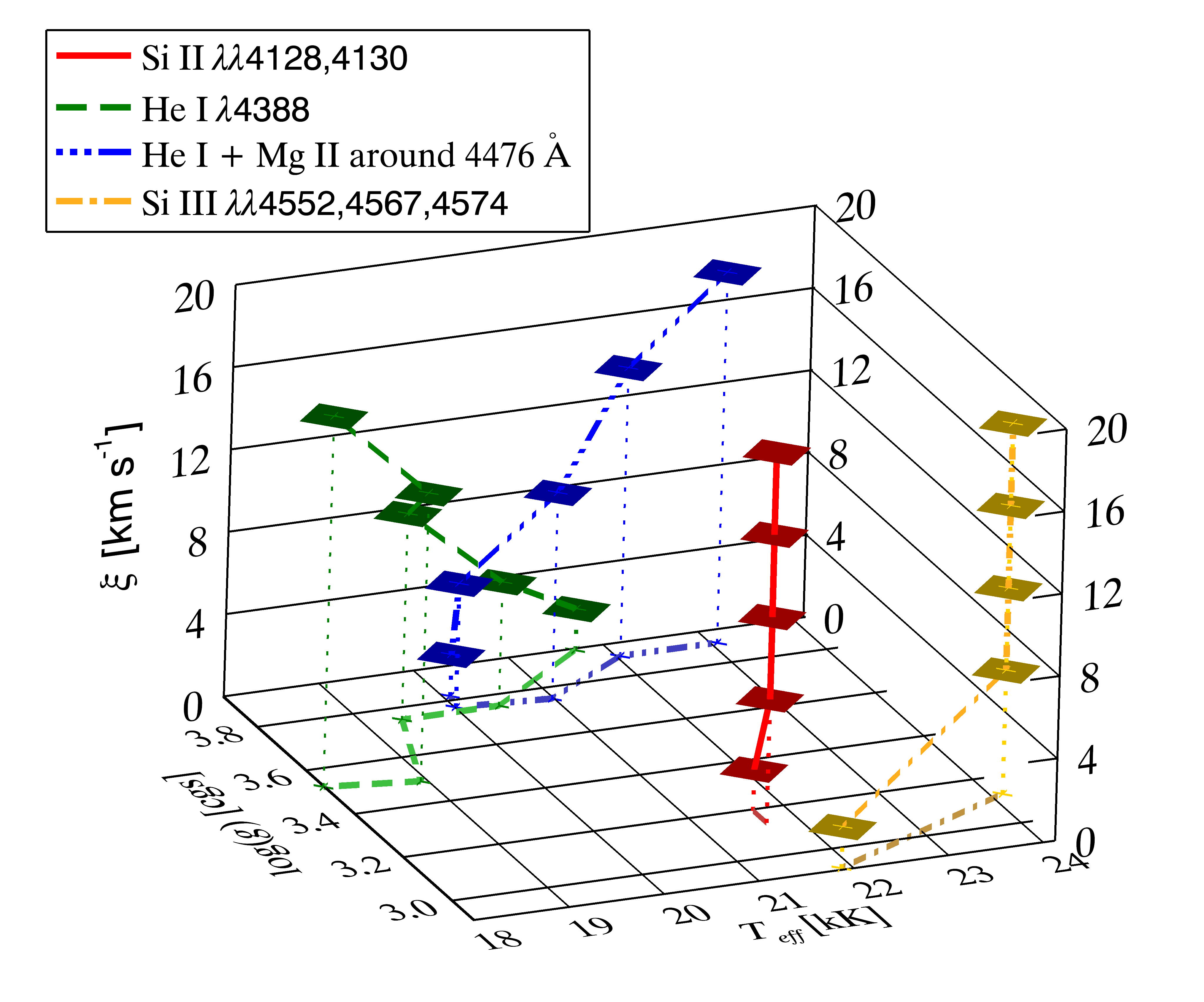}
  \caption{The most probable pair [$T_{eff}$, $\log(g)$] as a function of $\xi$
for different lines (as indicated in the plot) for the same artificial spectrum as in Figure \ref{fig:HbetHgam}.
The behaviour as a function of $\xi$ is shown in the plane [$T_{eff}$, $\log(g)$]. 
The likelihoods are marginalized over all $v\,\sin(i)$ values. } 
\label{fig:vtuinfl}
\end{figure}

Fixing or constraining the microturbulence velocity with only a few lines can lead to large errors for the other parameter estimates
as well. Figure \ref{fig:vtuinfl} shows an example of the most probable pairs [$T_{eff}$, log($g$)] for $\xi=2$, 6, 10, 14, and $18$~km~s$^{-1}$ 
when calculating the likelihoods with the lines He\,{\tt I}\,$\lambda$4388, He\,{\tt I}+Mg\,{\tt II} around 4476~\r{A}, 
Si\,{\tt II}\,$\lambda\lambda$4128,4130,  and Si\,{\tt III}\,$\lambda\lambda$4552,4567,4574 of the artificial spectrum with the following
specifications: $T_{eff}=21350$~K, $\log(g)=3.8$, $\xi=3$~km~s$^{-1}$, and $v\,\sin(i)=15$~km~s$^{-1}$. 
This shows that each line 
has a different sensitivity to the microturbulence velocity and, depending on the line or the group of lines chosen as a diagnostic, 
the result differs in $T_{eff}$ and $\log(g)$ but also in $\xi$. For instance here, the most probable value of $\xi$ given by the helium lines
is $10$~km~s$^{-1}$ (when marginalizing over all the other parameters) while the silicon lines give $2$~km~s$^{-1}$.
Thus, it is very important to consider $\xi$ as a free parameter as is usually done with $T_{eff}$, $\log(g)$, and $v\,\sin(i)$, and 
to constrain these four parameters simultaneously with as many lines as possible. Also it is even more important to consider 
the microturbulence velocity as a free parameter when studying the abundance, since it has a significant impact on this parameter as 
well \citep{Nieva2010}. \\

\section{Conclusions}

We described here a spectral analysis method based on Bayesian statistics that simultaneously constrains four stellar
parameters (effective temperature, surface gravity, projected rotational velocity, and microturbulence velocity). The method cumulates the information
provided simultaneously by all available lines in a given spectrum. This allow us to find the best global solution for all the lines and all the parameters 
at the same time, while providing reduced uncertainties. The Bayesian formalism naturally gives the uncertainties for each parameter
in relation with the other parameter possible values, with the data uncertainties, and with the limitations of the models used.

This method is completely objective since it does not rely on the user's visual judgement, nor does it depend on starting-parameter 
estimates. We have shown that adopting an inaccurate starting value or fixing the value of a given parameter, along with using a classical 
iterative method, can lead to substantial errors on the other parameters. Also, because our method considers all the parameters
and all the lines at the same time, it does not depend on the sensitivity of a specific diagnostic line.

This method is also self-consistent and efficient even when it is applied to spectra with high noise level or when underestimating
the quality of the data. Moreover, since our method cumulates the constraints of all available lines, it is still able to give 
accurate results with small uncertainties even though nearly all the lines are heavily blended due to the modest
resolution (2.3$\,$\r{A}) of our data. 

Note also, that the method actually works with atmospheric models from TLUSTY but also from ATLAS9 and PHOENIX \citep{2010ascl.soft10056B}, 
and can be easily adapted to virtually any atmospheric model. Furthermore, it is a fast method, once the basic preparation is done
(preparation of the data as suitable input for the code and creation of the basic grids) the complete basic parameters analysis 
of a star takes less than 3 minutes on a typical portable computer (we use a Pentium(R) Dual-Core CPU T4300 @ 2.10GHz with 4 Go of RAM).

Finally, the comparison of our results with those from the literature is very satisfying overall. Most of the differences found are
easily explained in terms of the method used (an iterative method versus our simultaneous method) and the number and choice of the 
diagnostic lines used (few lines for specific parameters versus, as we did, all available lines with the same weight for all the parameters).
One important behaviour is pointed out in this study: the microturbulence velocity is often underestimated for the B dwarfs as well as for 
the giants and supergiants. Furthermore, we also confirmed that cluster B stars are on average faster rotators than field B stars.

Now that we have demonstrated that we can successfully apply our method to gather stellar parameters using solar abundances, 
we will investigate in a future paper the impact and efficiency of our method for the abundance determination of various chemical elements.

\section*{Acknowledgments}

We thank Ivan Hubeny for generously providing us with the latest version of the SYNPLOT package and for helpful suggestions and comments.
 We also thank Anthony Moffat and the anonymous referee for a critical reading of the original version of this paper. 
This work was supported by the Natural Sciences and Engineering Research Council of Canada and
by the Fonds Qu\'eb\'ecois de la Recherche sur la Nature et les Technologies of the Government
of Qu\'ebec.

\bibliographystyle{mn2e}
\bibliography{biblio}

\clearpage
\appendix
\onecolumn

\section{Supplementary Figures}

\begin{figure}
\includegraphics[scale=0.12]{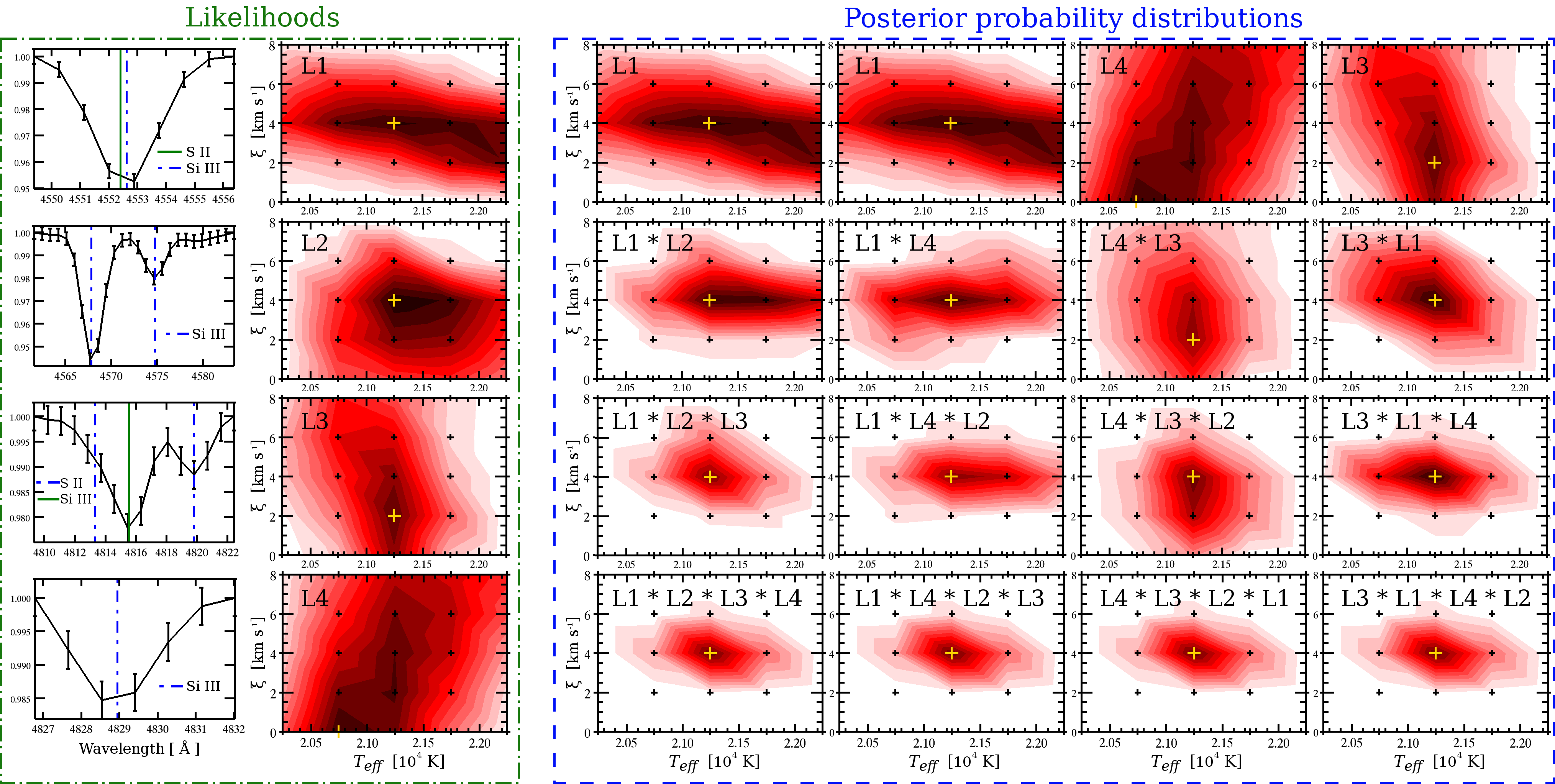}
\caption{ 
Illustration of the commutativity of the analysis. The lines from Fig \ref{fig:methex} are used here with a similar grid of synthetic spectra
where all the parameters other than $T_{eff}$ and $\xi$ have been marginalized. The green dash-dotted frame on the left shows each line 
considered with its respective likelihood, while the blue dashed frames holds the posterior probability distribution obtained when 
considering several possible combinations of the likelihoods. From the top to the bottom, each column of the blue frame shows the 
evolution of the posterior probability distribution as new lines are added to the analysis. While the order in which the lines are added 
differs from column to column, the final posterior probability distribution (bottom row) remains the same in each case. Note that the 
products written in each graphic (such as L1*L2*L3) are simplified for clarity. In fact, each posterior distribution is the normalised
 product of the likelihoods as stated in equation \ref{FBT}.}
\label{fig:comex}
\end{figure}

\newpage

\section[]{Tables of stellar Parameters}

\begin{center}
\begin{longtable}{cclll@{}l@{}c}
\caption[stellar parameter field]{Basic Parameters of Field B Stars} \label{tab:pftab} \\

\hline \multicolumn{1}{c}{\textbf{}} &
\multicolumn{1}{c}{\textbf{Star Id.}} &
\multicolumn{1}{c}{\textbf{$T_{eff}$}} & 
\multicolumn{1}{c}{\textbf{$\log(g)$}} &
\multicolumn{1}{c}{\textbf{$v\,\sin(i)$}} &
\multicolumn{1}{c}{\textbf{$\xi$}} &
\multicolumn{1}{c}{\textbf{ref}} \\
\multicolumn{1}{c}{\textbf{}} &
\multicolumn{1}{c}{\textbf{}} &
\multicolumn{1}{c}{\textbf{(K)}} & 
\multicolumn{1}{c}{\textbf{}} &
\multicolumn{1}{c}{\textbf{(km s$^{-1}$)}} &
\multicolumn{1}{c}{\textbf{(km s$^{-1}$)}} &
\multicolumn{1}{c}{\textbf{}} \\ \hline \\
\endfirsthead

\multicolumn{7}{c}%
{{Table \thetable{} -- continued from previous page}} \\
\hline \multicolumn{1}{c}{\textbf{}} &
\multicolumn{1}{c}{\textbf{Star Id.}} &
\multicolumn{1}{c}{\textbf{$T_{eff}$}} & 
\multicolumn{1}{c}{\textbf{$\log(g)$}} &
\multicolumn{1}{c}{\textbf{$v\,\sin(i)$}} &
\multicolumn{1}{c}{\textbf{$\xi$}} &
\multicolumn{1}{c}{\textbf{ref}} \\
\multicolumn{1}{c}{\textbf{}} &
\multicolumn{1}{c}{\textbf{}} &
\multicolumn{1}{c}{\textbf{(K)}} & 
\multicolumn{1}{c}{\textbf{}} &
\multicolumn{1}{c}{\textbf{(km s$^{-1}$)}} &
\multicolumn{1}{c}{\textbf{(km s$^{-1}$)}} &
\multicolumn{1}{c}{\textbf{}} \\ \hline \\
\endhead

\hline \multicolumn{7}{r}{{Continued on next page}} \\ \hline
\endfoot

\hline \\ \hline \multicolumn{7}{l}{{1: Spectrum with an effective resolution of 0.9~\r{A}.}}\\
\multicolumn{7}{l}{{References: TW = This Work using models from TLUSTY with the solar composition from \citet{Asplund2009};}}\\
\multicolumn{7}{l}{{HG10 = \citet{Huang2010}; LEF10 = \citet{Lefever2010}; TAK10 = \citet{Takeda2010};}} \\ 
\multicolumn{7}{l}{{SEA08 = \citet{Searle2008}; MAR08 = \citet{Markova2008}; DAF07 = \citet{Daflon2007};}}\\
\multicolumn{7}{l}{{HG06 = \citet{Huang2006a,Huang2006b}; LYU05 = \citet{Lyubimkov2005}; and AND99 = \citet{Andrievsky1999}.}}\\
\endlastfoot
 
& HD886 & 21250 $\pm$ 130 & 3.98 $\pm$ 0.04 & \phantom{11}31 $^{+9}_{-11}$ & \phantom{1111}8 $\pm$ 1 & TW \\*
& & \textit{21600 $\pm$ 1000} & \textit{3.7 $\pm$ 0.3} & \phantom{111}\textit{8 $\pm$ 2} & \phantom{1111}\textit{3.3 } & \textit{AND99} \\*
& & \textit{23000 $\pm$ 1000} & \textit{3.8 $\pm$ 0.1} & \phantom{11}\textit{10 $\pm$ 1} & \phantom{1111}\textit{1 $^{+2}_{-1}$} & \textit{LEF10} \\*
& & \textit{21667 $\pm$ 650} & \textit{3.8 $\pm$ 0.2} & \phantom{111}\textit{0 $\pm$ 9} & \phantom{1111}\textit{3 $\pm$ 2} & \textit{TAK10} \\[5pt]
& HD3827 & 29440 $^{+170}_{-240}$ & 4.27 $\pm$ 0.04 & \phantom{1}155 $\pm$ 5 & \phantom{1111}6 $\pm$ 1 & TW \\*
& & \textit{24678 $\pm$ 450} & \textit{3.836 $\pm$ 0.038} & \phantom{1}\textit{164 $\pm$ 10} & \phantom{1111}\textit{2 } & \textit{HG10} \\[5pt]
& HD5882 & 18250 $\pm$ 130 & 4.27 $\pm$ 0.04 & \phantom{1}330 $^{+5}_{-10}$ & \phantom{1111}7 $^{+1}_{-2}$ & TW \\*
& & \textit{15626 $\pm$ 150} & \textit{3.731 $\pm$ 0.021} & \phantom{1}\textit{358 $\pm$ 5} & \phantom{1111}\textit{2 } & \textit{HG10} \\
\end{longtable}
\end{center}

\newpage

\begin{center}
\begin{longtable}{cllll@{}l@{}c}
\caption[stellar Parameters Cluster]{Basic Parameters of Cluster B Stars} \label{tab:pftabcl} \\

\hline \multicolumn{1}{c}{\textbf{Cluster Id.}} &
\multicolumn{1}{c}{\textbf{Star Id.}} &
\multicolumn{1}{c}{\textbf{$T_{eff}$}} & 
\multicolumn{1}{c}{\textbf{$\log(g)$}} &
\multicolumn{1}{c}{\textbf{$v\,\sin(i)$}} &
\multicolumn{1}{c}{\textbf{$\xi$}} &
\multicolumn{1}{c}{\textbf{ref}} \\
\multicolumn{1}{c}{\textbf{}} &
\multicolumn{1}{c}{\textbf{}} &
\multicolumn{1}{c}{\textbf{(K)}} & 
\multicolumn{1}{c}{\textbf{}} &
\multicolumn{1}{c}{\textbf{(km s$^{-1}$)}} &
\multicolumn{1}{c}{\textbf{(km s$^{-1}$)}} &
\multicolumn{1}{c}{\textbf{}} \\ \hline \\
\endfirsthead

\multicolumn{7}{c}%
{{Table \thetable{} -- continued from previous page}} \\
\hline \multicolumn{1}{c}{\textbf{Cluster Id.}} &
\multicolumn{1}{c}{\textbf{Star Id.}} &
\multicolumn{1}{c}{\textbf{$T_{eff}$}} & 
\multicolumn{1}{c}{\textbf{$\log(g)$}} &
\multicolumn{1}{c}{\textbf{$v\,\sin(i)$}} &
\multicolumn{1}{c}{\textbf{$\xi$}} &
\multicolumn{1}{c}{\textbf{ref}} \\
\multicolumn{1}{c}{\textbf{}} &
\multicolumn{1}{c}{\textbf{}} &
\multicolumn{1}{c}{\textbf{(K)}} & 
\multicolumn{1}{c}{\textbf{}} &
\multicolumn{1}{c}{\textbf{(km s$^{-1}$)}} &
\multicolumn{1}{c}{\textbf{(km s$^{-1}$)}} &
\multicolumn{1}{c}{\textbf{}} \\ \hline \\
\endhead

\hline \multicolumn{7}{r}{{Continued on next page}} \\ \hline
\endfoot

\hline \\ \hline
\multicolumn{7}{l}{{References: TW = This Work using models from TLUSTY with the solar composition from \citet{Asplund2009}; }}\\
\multicolumn{7}{l}{{HG10 = \citet{Huang2010}; and HG06 = \citet{Huang2006a,Huang2006b}. }} \\ 
\endlastfoot

NGC1960 & BD+34 1103s & 19660 $^{+220}_{-160}$ & 4.30 $\pm$ 0.04 & \phantom{1}288 $^{+9}_{-6}$ & \phantom{1111}4 $^{+1}_{-2}$ & TW \\*
\textit{} & \textit{} & \textit{17297 $\pm$ 150} & \textit{3.983 $\pm$ 0.015} & \phantom{11}\textit{77 $\pm$ 7} & \phantom{1111}\textit{2 } & \textit{HG10} \\[4pt]
& SAB20 & 15000 $^{+250}_{-0}$ & 4.20 $^{+0.04}_{-0.08}$ & \phantom{1}301 $^{+11}_{-9}$ & \phantom{1111}6 $\pm$ 3 & TW \\*
\textit{} & \textit{} & \textit{14845 $\pm$ 100} & \textit{4.242 $\pm$ 0.018} & \phantom{1}\textit{141 $\pm$ 7} & \phantom{1111}\textit{2 } & \textit{HG10} \\[4pt]
& SAB22 & 15000 $^{+130}_{-0}$ & 4.35 $\pm$ 0.04 & \phantom{1}118 $^{+7}_{-8}$ & \phantom{1111}3 $^{+2}_{-1}$ & TW \\*
\textit{} & \textit{} & \textit{13579 $\pm$ 100} & \textit{4.071 $\pm$ 0.019} & \phantom{1}\textit{135 $\pm$ 6} & \phantom{1111}\textit{2 } & \textit{HG10} \\[4pt]
& SAB23 & 15000 $^{+250}_{-0}$ & 4.5 $^{+0}_{-0.06}$ & \phantom{1}300 $\pm$ 20 & \phantom{1111}0 $^{+4}_{-0}$ & TW \\*
\textit{} & \textit{} & \textit{13715 $\pm$ 150} & \textit{4.267 $\pm$ 0.028} & \phantom{1}\textit{302 $\pm$ 11} & \phantom{1111}\textit{2 } & \textit{HG10} \\
\end{longtable}

\end{center}

\newpage

\begin{center}
\begin{longtable}{clll@{}l@{}ccc}
\caption[Stellar parameter field lefevre]{Basic Parameters of Field B Stars with Modified He and Si Abundances from \citet{Lefever2010}} \label{tab:lef_tab} \\

\hline \\
\multicolumn{1}{c}{\textbf{Star Id.}} &
\multicolumn{1}{c}{\textbf{$T_{eff}$}} & 
\multicolumn{1}{c}{\textbf{$\log(g)$}} &
\multicolumn{1}{c}{\textbf{$v\,\sin(i)$}} &
\multicolumn{1}{c}{\textbf{$\xi$}} &
\multicolumn{1}{c}{\textbf{$\log(\frac{He}{H})+12$}} &
\multicolumn{1}{c}{\textbf{$\log(\frac{Si}{H})+12$}} &
\multicolumn{1}{c}{\textbf{ref}} \\
\multicolumn{1}{c}{\textbf{}} &
\multicolumn{1}{c}{\textbf{(K)}} & 
\multicolumn{1}{c}{\textbf{}} &
\multicolumn{1}{c}{\textbf{(km s$^{-1}$)}} &
\multicolumn{1}{c}{\textbf{(km s$^{-1}$)}} &
\multicolumn{1}{c}{\textbf{}} &
\multicolumn{1}{c}{\textbf{}} &
\multicolumn{1}{c}{\textbf{}} \\ \hline \\
\endfirsthead

\multicolumn{8}{c}%
{{Table \thetable{} -- continued from previous page}} \\
\hline \\
\multicolumn{1}{c}{\textbf{Star Id.}} &
\multicolumn{1}{c}{\textbf{$T_{eff}$}} & 
\multicolumn{1}{c}{\textbf{$\log(g)$}} &
\multicolumn{1}{c}{\textbf{$v\,\sin(i)$}} &
\multicolumn{1}{c}{\textbf{$\xi$}} &
\multicolumn{1}{c}{\textbf{$\log(\frac{He}{H})+12$}} &
\multicolumn{1}{c}{\textbf{$\log(\frac{Si}{H})+12$}} &
\multicolumn{1}{c}{\textbf{ref}} \\
\multicolumn{1}{c}{\textbf{}} &
\multicolumn{1}{c}{\textbf{(K)}} & 
\multicolumn{1}{c}{\textbf{}} &
\multicolumn{1}{c}{\textbf{(km s$^{-1}$)}} &
\multicolumn{1}{c}{\textbf{(km s$^{-1}$)}} &
\multicolumn{1}{c}{\textbf{}} &
\multicolumn{1}{c}{\textbf{}} &
\multicolumn{1}{c}{\textbf{}} \\ \hline \\
\endhead

\hline \multicolumn{8}{r}{{Continued on next page}} \\ \hline
\endfoot

\hline \\ \hline \multicolumn{8}{l}{{References: TW = This Work; LEF10 = \citet{Lefever2010}.}}\\
\endlastfoot
HD160762 & 18090 $^{+130}_{-120}$ & 3.97 $\pm$ 0.04 & \phantom{111}3 $^{+7}_{-3}$ & \phantom{1111}9 $\pm$ 1 & 11 & 7.14 & TW \\*
& \textit{19500 $\pm$ 500} & \textit{4.1 $\pm$ 0.1} & \phantom{111}\textit{8 $\pm$ 2} & \phantom{1111}\textit{3 $\pm$ 2} &  &  & \textit{LEF10} \\[4pt]
HD44700 & 17250 $\pm$ 130 & 4.22 $\pm$ 0.04 & \phantom{111}8 $^{+16}_{-8}$ & \phantom{1111}6 $\pm$ 1 & 11 & 7.07 & TW \\*
& \textit{17500 $\pm$ 500} & \textit{3.8 $\pm$ 0.1} & \phantom{111}\textit{8 $\pm$ 1} & \phantom{1111}\textit{6 $\pm$ 2} &  &  & \textit{LEF10} \\[4pt]
HD45418 & 16750 $\pm$ 130 & 4.27 $\pm$ 0.04 & \phantom{1}247 $^{+8}_{-7}$ & \phantom{1111}6 $\pm$ 1 & 11 & 7.51 & TW \\*
& \textit{16000 $\pm$ 500} & \textit{4. $\pm$ 0.2} & \phantom{1}\textit{237 $\pm$ 28} & \phantom{1111}\textit{3 $\pm$ 2} &  &  & \textit{LEF10} \\[4pt]
HD886 & 21120 $^{+140}_{-240}$ & 3.95 $\pm$ 0.04 & \phantom{11}34 $\pm$ 8 & \phantom{1111}9 $\pm$ 1 & 11 & 7.16 & TW \\*
& \textit{23000 $\pm$ 1000} & \textit{3.8 $\pm$ 0.1} & \phantom{11}\textit{10 $\pm$ 1} & \phantom{1111}\textit{1 $^{+2}_{-1}$} &  &  & \textit{LEF10} \\
\end{longtable}

\end{center}

\newpage

\begin{center}
\begin{longtable}{clll@{}l@{}ccc}
\caption[Basic parameter field takeda]{Basic Parameters of Field B Stars with Modified O and Ne Abundances from \citet{Takeda2010}} \label{tab:tak_tab} \\

\hline \\
\multicolumn{1}{c}{\textbf{Star Id.}} &
\multicolumn{1}{c}{\textbf{$T_{eff}$}} & 
\multicolumn{1}{c}{\textbf{$\log(g)$}} &
\multicolumn{1}{c}{\textbf{$v\,\sin(i)$}} &
\multicolumn{1}{c}{\textbf{$\xi$}} &
\multicolumn{1}{c}{\textbf{$\log(\frac{O}{H})+12$}} &
\multicolumn{1}{c}{\textbf{$\log(\frac{Ne}{H})+12$}} &
\multicolumn{1}{c}{\textbf{ref}} \\
\multicolumn{1}{c}{\textbf{}} &
\multicolumn{1}{c}{\textbf{(K)}} & 
\multicolumn{1}{c}{\textbf{}} &
\multicolumn{1}{c}{\textbf{(km s$^{-1}$)}} &
\multicolumn{1}{c}{\textbf{(km s$^{-1}$)}} &
\multicolumn{1}{c}{\textbf{}} &
\multicolumn{1}{c}{\textbf{}} &
\multicolumn{1}{c}{\textbf{}} \\ \hline \\
\endfirsthead

\multicolumn{8}{c}%
{{Table \thetable{} -- continued from previous page}} \\
\hline \\
\multicolumn{1}{c}{\textbf{Star Id.}} &
\multicolumn{1}{c}{\textbf{$T_{eff}$}} & 
\multicolumn{1}{c}{\textbf{$\log(g)$}} &
\multicolumn{1}{c}{\textbf{$v\,\sin(i)$}} &
\multicolumn{1}{c}{\textbf{$\xi$}} &
\multicolumn{1}{c}{\textbf{$\log(\frac{O}{H})+12$}} &
\multicolumn{1}{c}{\textbf{$\log(\frac{Ne}{H})+12$}} &
\multicolumn{1}{c}{\textbf{ref}} \\
\multicolumn{1}{c}{\textbf{}} &
\multicolumn{1}{c}{\textbf{(K)}} & 
\multicolumn{1}{c}{\textbf{}} &
\multicolumn{1}{c}{\textbf{(km s$^{-1}$)}} &
\multicolumn{1}{c}{\textbf{(km s$^{-1}$)}} &
\multicolumn{1}{c}{\textbf{}} &
\multicolumn{1}{c}{\textbf{}} &
\multicolumn{1}{c}{\textbf{}} \\ \hline \\
\endhead

\hline \multicolumn{8}{r}{{Continued on next page}} \\ \hline
\endfoot

\hline \\ \hline \multicolumn{8}{l}{{References: TW = This Work; TAK10 = \citet{Takeda2010}.}}\\
\endlastfoot
HD160762 & 16880 $\pm$ 130 & 3.81 $\pm$ 0.04 & \phantom{111}3 $^{+12}_{-3}$ & \phantom{1111}8 $^{+1}_{-2}$ & 8.74 & 8.08 & TW \\*
& \textit{17440 $\pm$ 523} & \textit{3.9 $\pm$ 0.2} & \phantom{111}\textit{0 $\pm$ 9} & \phantom{1111}\textit{3 $\pm$ 2} &  &  & \textit{TAK10} \\[4pt]
HD184171 & 17000 $\pm$ 130 & 3.86 $\pm$ 0.04 & \phantom{111}3 $^{+7}_{-3}$ & \phantom{1111}3 $\pm$ 1 & 8.69 & 8.01 & TW \\*
& \textit{15858 $\pm$ 476} & \textit{3.5 $\pm$ 0.2} & \phantom{11}\textit{15 $\pm$ 9} & \phantom{1111}\textit{3 $\pm$ 2} &  &  & \textit{TAK10} \\[4pt]
HD185330 & 15000 $^{+250}_{-0}$ & 3.70 $\pm$ 0.04 & \phantom{1}120 $\pm$ 12 & \phantom{1111}2 $^{+2}_{-1}$ & 8.35 & 8.15 & TW \\*
& \textit{16167 $\pm$ 485} & \textit{3.7 $\pm$ 0.2} & \phantom{11}\textit{15 $\pm$ 9} & \phantom{1111}\textit{3 $\pm$ 2} &  &  & \textit{TAK10} \\[4pt]
HD188892 & 15120 $^{+130}_{-120}$ & 3.57 $\pm$ 0.04 & \phantom{111}3 $^{+7}_{-3}$ & \phantom{1111}4 $\pm$ 1 & 8.67 & 8.01 & TW \\*
& \textit{14008 $\pm$ 420} & \textit{3.3 $\pm$ 0.2} & \phantom{11}\textit{30 $\pm$ 9} & \phantom{1111}\textit{3 $\pm$ 2} &  &  & \textit{TAK10} \\
\end{longtable}

\end{center}

\newpage

\begin{center}
\begin{longtable}{clll@{}l@{}cccc}
\caption[Basic parameter field Searle]{Basic Parameters of Field B Stars with Modified C, N, O Abundances from \citet{Searle2008}} \label{tab:Sear_tab} \\

\hline \\
\multicolumn{1}{c}{\textbf{Star Id.}} &
\multicolumn{1}{c}{\textbf{$T_{eff}$}} & 
\multicolumn{1}{c}{\textbf{$\log(g)$}} &
\multicolumn{1}{c}{\textbf{$v\,\sin(i)$}} &
\multicolumn{1}{c}{\textbf{$\xi$}} &
\multicolumn{1}{c}{\textbf{$\log(\frac{C}{H})+12$}} &
\multicolumn{1}{c}{\textbf{$\log(\frac{N}{H})+12$}} &
\multicolumn{1}{c}{\textbf{$\log(\frac{O}{H})+12$}} &
\multicolumn{1}{c}{\textbf{ref}} \\
\multicolumn{1}{c}{\textbf{}} &
\multicolumn{1}{c}{\textbf{(K)}} & 
\multicolumn{1}{c}{\textbf{}} &
\multicolumn{1}{c}{\textbf{(km s$^{-1}$)}} &
\multicolumn{1}{c}{\textbf{(km s$^{-1}$)}} &
\multicolumn{1}{c}{\textbf{}} &
\multicolumn{1}{c}{\textbf{}} &
\multicolumn{1}{c}{\textbf{}} &
\multicolumn{1}{c}{\textbf{}} \\ \hline \\
\endfirsthead

\multicolumn{9}{c}%
{{Table \thetable{} -- continued from previous page}} \\
\hline \\
\multicolumn{1}{c}{\textbf{Star Id.}} &
\multicolumn{1}{c}{\textbf{$T_{eff}$}} & 
\multicolumn{1}{c}{\textbf{$\log(g)$}} &
\multicolumn{1}{c}{\textbf{$v\,\sin(i)$}} &
\multicolumn{1}{c}{\textbf{$\xi$}} &
\multicolumn{1}{c}{\textbf{$\log(\frac{C}{H})+12$}} &
\multicolumn{1}{c}{\textbf{$\log(\frac{N}{H})+12$}} &
\multicolumn{1}{c}{\textbf{$\log(\frac{O}{H})+12$}} &
\multicolumn{1}{c}{\textbf{ref}} \\
\multicolumn{1}{c}{\textbf{}} &
\multicolumn{1}{c}{\textbf{(K)}} & 
\multicolumn{1}{c}{\textbf{}} &
\multicolumn{1}{c}{\textbf{(km s$^{-1}$)}} &
\multicolumn{1}{c}{\textbf{(km s$^{-1}$)}} &
\multicolumn{1}{c}{\textbf{}} &
\multicolumn{1}{c}{\textbf{}} &
\multicolumn{1}{c}{\textbf{}} &
\multicolumn{1}{c}{\textbf{}} \\ \hline \\
\endhead

\hline \multicolumn{9}{r}{{Continued on next page}} \\ \hline
\endfoot

\hline \\ \hline \multicolumn{9}{l}{{References: TW = This Work; SEA08 = \citet{Searle2008}.}}\\
\endlastfoot
HD190066 & 23750 $\pm$ 190 & 2.90 $\pm$ 0.04 & \phantom{11}95 $\pm$ 9 & \phantom{111}25 $\pm$ 1 & 7.88 & 8.15 & 8.53 & TW \\*
& \textit{21000 $\pm$ 1000} & \textit{2.88 $\pm$ 0.25} & \phantom{11}\textit{82 $\pm$ 20} & \phantom{111}\textit{15 } &  &  &  & \textit{SEA08} \\[4pt]
HD191243 & 15120 $^{+130}_{-120}$ & 2.70 $\pm$ 0.04 & \phantom{111}3 $^{+7}_{-3}$ & \phantom{111}11 $\pm$ 1 & 7.70 & 7.65 & & TW \\*
& \textit{14500 $\pm$ 1000} & \textit{2.75 $\pm$ 0.25} & \phantom{11}\textit{38 $\pm$ 20} & \phantom{111}\textit{20 } &  &  &  & \textit{SEA08} \\[4pt]
HD192660 & 25140 $\pm$ 160 & 2.88 $\pm$ 0.03 & \phantom{1}140 $\pm$ 5 & \phantom{111}29 $^{+1}_{-2}$ & 8.02 & 7.51 & 8.73 & TW \\*
& \textit{30000 $\pm$ 1000} & \textit{3.25 $\pm$ 0.25} & \phantom{11}\textit{94 $\pm$ 20} & \phantom{111}\textit{20 } &  &  &  & \textit{SEA08} \\[4pt]
HD193183 & 19610 $^{+140}_{-240}$ & 2.65 $\pm$ 0.04 & \phantom{11}32 $\pm$ 12 & \phantom{111}23 $\pm$ 1 & 7.66 & 8.15 & 8.73 & TW \\*
& \textit{18500 $\pm$ 1000} & \textit{2.63 $\pm$ 0.25} & \phantom{11}\textit{68 $\pm$ 20} & \phantom{111}\textit{20 } &  &  &  & \textit{SEA08} \\
\end{longtable}

\end{center}

\begin{center}
\begin{longtable}{clll@{}l@{}ccc}
\caption[Basic parameter field markova]{Basic Parameters of Field B Stars with Modified He and Si Abundances from \citet{Markova2008}} \label{tab:mark_tab} \\

\hline \\
\multicolumn{1}{c}{\textbf{Star Id.}} &
\multicolumn{1}{c}{\textbf{$T_{eff}$}} & 
\multicolumn{1}{c}{\textbf{$\log(g)$}} &
\multicolumn{1}{c}{\textbf{$v\,\sin(i)$}} &
\multicolumn{1}{c}{\textbf{$\xi$}} &
\multicolumn{1}{c}{\textbf{$\log(\frac{He}{H})+12$}} &
\multicolumn{1}{c}{\textbf{$\log(\frac{Si}{H})+12$}} &
\multicolumn{1}{c}{\textbf{ref}} \\
\multicolumn{1}{c}{\textbf{}} &
\multicolumn{1}{c}{\textbf{(K)}} & 
\multicolumn{1}{c}{\textbf{}} &
\multicolumn{1}{c}{\textbf{(km s$^{-1}$)}} &
\multicolumn{1}{c}{\textbf{(km s$^{-1}$)}} &
\multicolumn{1}{c}{\textbf{}} &
\multicolumn{1}{c}{\textbf{}} &
\multicolumn{1}{c}{\textbf{}} \\ \hline \\
\endfirsthead

\multicolumn{8}{c}%
{{Table \thetable{} -- continued from previous page}} \\
\hline \\
\multicolumn{1}{c}{\textbf{Star Id.}} &
\multicolumn{1}{c}{\textbf{$T_{eff}$}} & 
\multicolumn{1}{c}{\textbf{$\log(g)$}} &
\multicolumn{1}{c}{\textbf{$v\,\sin(i)$}} &
\multicolumn{1}{c}{\textbf{$\xi$}} &
\multicolumn{1}{c}{\textbf{$\log(\frac{He}{H})+12$}} &
\multicolumn{1}{c}{\textbf{$\log(\frac{Si}{H})+12$}} &
\multicolumn{1}{c}{\textbf{ref}} \\
\multicolumn{1}{c}{\textbf{}} &
\multicolumn{1}{c}{\textbf{(K)}} & 
\multicolumn{1}{c}{\textbf{}} &
\multicolumn{1}{c}{\textbf{(km s$^{-1}$)}} &
\multicolumn{1}{c}{\textbf{(km s$^{-1}$)}} &
\multicolumn{1}{c}{\textbf{}} &
\multicolumn{1}{c}{\textbf{}} &
\multicolumn{1}{c}{\textbf{}} \\ \hline \\
\endhead

\hline \multicolumn{8}{r}{{Continued on next page}} \\ \hline
\endfoot

\hline \\ \hline \multicolumn{8}{l}{{References: TW = This Work; MAR08 = \citet{Markova2008}.}}\\
\endlastfoot
HD191243 & 15240 $^{+140}_{-240}$ & 2.77 $^{+0.04}_{-0.08}$ & \phantom{111}3 $^{+7}_{-3}$ & \phantom{111}11 $\pm$ 1 & 10.954 & 7.48 & TW \\*
& \textit{14800 $\pm$ 1000} & \textit{2.6 $\pm$ 0.2} & \phantom{11}\textit{38 $\pm$ 10} & \phantom{1111}\textit{8 $\pm$ 5} &  &  & \textit{MAR08} \\[4pt]
HD198478 & 16270 $^{+230}_{-150}$ & 2.08 $^{+0.08}_{-0.04}$ & \phantom{111}2 $^{+28}_{-2}$ & \phantom{111}26 $^{+1}_{-2}$ & 11.176 & 7.58 & TW \\*
& \textit{17500 $\pm$ 1000} & \textit{2.1 $\pm$ 0.2} & \phantom{11}\textit{39 $\pm$ 10} & \phantom{1111}\textit{8 $\pm$ 5} &  &  & \textit{MAR08} \\[4pt]
HD206165 & 19380 $^{+120}_{-130}$ & 2.63 $\pm$ 0.04 & \phantom{11}76 $^{+9}_{-6}$ & \phantom{111}16 $^{+2}_{-1}$ & 11.176 & 7.58 & TW \\*
& \textit{19300 $\pm$ 1000} & \textit{2.5 $\pm$ 0.2} & \phantom{11}\textit{45 $\pm$ 10} & \phantom{1111}\textit{8 $\pm$ 5} &  &  & \textit{MAR08} \\
\end{longtable}

\end{center}

\newpage

\begin{center}
\begin{longtable}{clll@{}l@{}ccccccc}
\caption[Basic parameter field Daflon]{Basic Parameters of Field B Stars with Modified C, N, O, Mg, Al, and Si Abundances from \citet{Daflon2007}} \label{tab:daf_tab} \\

\hline \\
\multicolumn{1}{c}{\textbf{Star Id.}} &
\multicolumn{1}{c}{\textbf{$T_{eff}$}} & 
\multicolumn{1}{c}{\textbf{$\log(g)$}} &
\multicolumn{1}{c}{\textbf{$v\,\sin(i)$}} &
\multicolumn{1}{c}{\textbf{$\xi$}} &
\multicolumn{1}{c}{\textbf{$\epsilon(C)$}} &
\multicolumn{1}{c}{\textbf{$\epsilon(N)$}} &
\multicolumn{1}{c}{\textbf{$\epsilon(O)$}} &
\multicolumn{1}{c}{\textbf{$\epsilon(Mg)$}} &
\multicolumn{1}{c}{\textbf{$\epsilon(Al)$}} &
\multicolumn{1}{c}{\textbf{$\epsilon(Si)$}} &
\multicolumn{1}{c}{\textbf{ref}} \\
\multicolumn{1}{c}{\textbf{}} &
\multicolumn{1}{c}{\textbf{(K)}} & 
\multicolumn{1}{c}{\textbf{}} &
\multicolumn{1}{c}{\textbf{(km s$^{-1}$)}} &
\multicolumn{1}{c}{\textbf{(km s$^{-1}$)}} &
\multicolumn{1}{c}{\textbf{}} &
\multicolumn{1}{c}{\textbf{}} &
\multicolumn{1}{c}{\textbf{}} &
\multicolumn{1}{c}{\textbf{}} &
\multicolumn{1}{c}{\textbf{}} &
\multicolumn{1}{c}{\textbf{}} &
\multicolumn{1}{c}{\textbf{}} \\ \hline \\
\endfirsthead

\multicolumn{12}{c}%
{{Table \thetable{} -- continued from previous page}} \\
\hline \\
\multicolumn{1}{c}{\textbf{Star Id.}} &
\multicolumn{1}{c}{\textbf{$T_{eff}$}} & 
\multicolumn{1}{c}{\textbf{$\log(g)$}} &
\multicolumn{1}{c}{\textbf{$v\,\sin(i)$}} &
\multicolumn{1}{c}{\textbf{$\xi$}} &
\multicolumn{1}{c}{\textbf{$\epsilon(C)$}} &
\multicolumn{1}{c}{\textbf{$\epsilon(N)$}} &
\multicolumn{1}{c}{\textbf{$\epsilon(O)$}} &
\multicolumn{1}{c}{\textbf{$\epsilon(Mg)$}} &
\multicolumn{1}{c}{\textbf{$\epsilon(Al)$}} &
\multicolumn{1}{c}{\textbf{$\epsilon(Si)$}} &
\multicolumn{1}{c}{\textbf{ref}} \\
\multicolumn{1}{c}{\textbf{}} &
\multicolumn{1}{c}{\textbf{(K)}} & 
\multicolumn{1}{c}{\textbf{}} &
\multicolumn{1}{c}{\textbf{(km s$^{-1}$)}} &
\multicolumn{1}{c}{\textbf{(km s$^{-1}$)}} &
\multicolumn{1}{c}{\textbf{}} &
\multicolumn{1}{c}{\textbf{}} &
\multicolumn{1}{c}{\textbf{}} &
\multicolumn{1}{c}{\textbf{}} &
\multicolumn{1}{c}{\textbf{}} &
\multicolumn{1}{c}{\textbf{}} &
\multicolumn{1}{c}{\textbf{}} \\ \hline \\
\endhead

\hline \multicolumn{12}{r}{{Continued on next page}} \\ \hline
\endfoot

\hline \\ \hline \multicolumn{12}{l}{{References: TW = This Work; AND99 = \citet{Daflon2007}. Here, $\epsilon(X)=\log(X/H)+12$.}}\\
\endlastfoot
hd202347 & 23730 $^{+240}_{-260}$ & 4.27 $^{+0.04}_{-0.08}$ & \phantom{11}96 $^{+9}_{-6}$ & \phantom{1111}9 $\pm$ 1 &  & 7.54 & 8.54 & 7.35 & 6.28 & 7.13 & TW \\*
& \textit{23280 $\pm$ 900} & \textit{4.1 $\pm$ 0.1} & \phantom{1}\textit{119 $\pm$ 6} & \phantom{1111}\textit{6 $^{+1}_{-1.5}$} &  &  &  &  &  &  & \textit{DAF07} \\[4pt]
hd205794 & 25010 $^{+120}_{-130}$ & 3.95 $\pm$ 0.04 & \phantom{111}3 $^{+17}_{-3}$ & \phantom{111}11 $\pm$ 1 & 8.09 & 7.57 & 8.63 &  &  & 7.09 & TW \\*
& \textit{26890 $\pm$ 1000} & \textit{4.2 $\pm$ 0.1} & \phantom{11}\textit{14 $\pm$ 3} & \phantom{1111}\textit{8 $^{+1}_{-1.5}$} &  &  &  &  &  &  & \textit{DAF07} \\
\end{longtable}

\end{center}

\begin{center}
\begin{longtable}{clll@{}l@{}ccc}
\caption[Basic parameter field lyubimkov]{Basic Parameters of Field B Stars with Modified He and Mg Abundances from \citet{Lyubimkov2005}} \label{tab:Lyu_tab} \\

\hline \\
\multicolumn{1}{c}{\textbf{Star Id.}} &
\multicolumn{1}{c}{\textbf{$T_{eff}$}} & 
\multicolumn{1}{c}{\textbf{$\log(g)$}} &
\multicolumn{1}{c}{\textbf{$v\,\sin(i)$}} &
\multicolumn{1}{c}{\textbf{$\xi$}} &
\multicolumn{1}{c}{\textbf{$\log(\frac{He}{H})+12$}} &
\multicolumn{1}{c}{\textbf{$\log(\frac{Mg}{H})+12$}} &
\multicolumn{1}{c}{\textbf{ref}} \\
\multicolumn{1}{c}{\textbf{}} &
\multicolumn{1}{c}{\textbf{(K)}} & 
\multicolumn{1}{c}{\textbf{}} &
\multicolumn{1}{c}{\textbf{(km s$^{-1}$)}} &
\multicolumn{1}{c}{\textbf{(km s$^{-1}$)}} &
\multicolumn{1}{c}{\textbf{}} &
\multicolumn{1}{c}{\textbf{}} &
\multicolumn{1}{c}{\textbf{}} \\ \hline \\
\endfirsthead

\multicolumn{8}{c}%
{{Table \thetable{} -- continued from previous page}} \\
\hline \\
\multicolumn{1}{c}{\textbf{Star Id.}} &
\multicolumn{1}{c}{\textbf{$T_{eff}$}} & 
\multicolumn{1}{c}{\textbf{$\log(g)$}} &
\multicolumn{1}{c}{\textbf{$v\,\sin(i)$}} &
\multicolumn{1}{c}{\textbf{$\xi$}} &
\multicolumn{1}{c}{\textbf{$\log(\frac{He}{H})+12$}} &
\multicolumn{1}{c}{\textbf{$\log(\frac{Mg}{H})+12$}} &
\multicolumn{1}{c}{\textbf{ref}} \\
\multicolumn{1}{c}{\textbf{}} &
\multicolumn{1}{c}{\textbf{(K)}} & 
\multicolumn{1}{c}{\textbf{}} &
\multicolumn{1}{c}{\textbf{(km s$^{-1}$)}} &
\multicolumn{1}{c}{\textbf{(km s$^{-1}$)}} &
\multicolumn{1}{c}{\textbf{}} &
\multicolumn{1}{c}{\textbf{}} &
\multicolumn{1}{c}{\textbf{}} \\ \hline \\
\endhead

\hline \multicolumn{8}{r}{{Continued on next page}} \\ \hline
\endfoot

\hline \\ \hline \multicolumn{8}{l}{{References: TW = This Work; LYU05 = \citet{Lyubimkov2005}.}}\\
\endlastfoot
HD160762 & 17280 $^{+120}_{-130}$ & 3.88 $\pm$ 0.04 & \phantom{111}3 $^{+7}_{-3}$ & \phantom{1111}7 $\pm$ 1 & 11.057 & 7.62 & TW \\*
& \textit{17000 $\pm$ 250} & \textit{3.77 $\pm$ 0.12} & \phantom{11}\textit{15 } & \phantom{1111}\textit{1.3 $^{+1}_{-1.3}$} &  &  & \textit{LYU05} \\[4pt]
HD166182 & 19150 $^{+230}_{-150}$ & 3.55 $^{+0.07}_{-0.05}$ & \phantom{111}0 $^{+10}_{-0}$ & \phantom{1111}9 $\pm$ 1 & 11.045 & 7.39 & TW \\*
& \textit{20000 $\pm$ 600} & \textit{3.54 $\pm$ 0.16} & \phantom{11}\textit{44 } & \phantom{1111}\textit{4.7 $^{+4}_{-4.7}$} &  &  & \textit{LYU05} \\[4pt]
HD184171 & 16880 $^{+120}_{-130}$ & 3.85 $\pm$ 0.04 & \phantom{111}3 $^{+7}_{-3}$ & \phantom{1111}2 $\pm$ 1 & 11.041 & 7.89 & TW \\*
& \textit{16100 $\pm$ 700} & \textit{3.62 $\pm$ 0.15} & \phantom{11}\textit{29 } & \phantom{1111}\textit{0 $\pm$ 0} &  &  & \textit{LYU05} \\[4pt]
HD196035 & 18210 $^{+170}_{-210}$ & 4.18 $\pm$ 0.04 & \phantom{111}3 $^{+7}_{-3}$ & \phantom{1111}6 $^{+2}_{-1}$ & 11.065 & 7.66 & TW \\*
& \textit{17100 $\pm$ 250} & \textit{4.0 $\pm$ 0.08} & \phantom{11}\textit{34 } & \phantom{1111}\textit{2 $\pm$ 2} &  &  & \textit{LYU05} \\
\end{longtable}

\end{center}

\newpage

\begin{center}
\begin{longtable}{clll@{}l@{}ccc}
\caption[Basic parameter field andreievsky]{Basic Parameters of Field B Stars with Modified C and N Abundances from \citet{Andrievsky1999}} \label{tab:and_tab} \\

\hline \\
\multicolumn{1}{c}{\textbf{Star Id.}} &
\multicolumn{1}{c}{\textbf{$T_{eff}$}} & 
\multicolumn{1}{c}{\textbf{$\log(g)$}} &
\multicolumn{1}{c}{\textbf{$v\,\sin(i)$}} &
\multicolumn{1}{c}{\textbf{$\xi$}} &
\multicolumn{1}{c}{\textbf{$\log(\frac{C}{H})+12$}} &
\multicolumn{1}{c}{\textbf{$\log(\frac{N}{H})+12$}} &
\multicolumn{1}{c}{\textbf{ref}} \\
\multicolumn{1}{c}{\textbf{}} &
\multicolumn{1}{c}{\textbf{(K)}} & 
\multicolumn{1}{c}{\textbf{}} &
\multicolumn{1}{c}{\textbf{(km s$^{-1}$)}} &
\multicolumn{1}{c}{\textbf{(km s$^{-1}$)}} &
\multicolumn{1}{c}{\textbf{}} &
\multicolumn{1}{c}{\textbf{}} &
\multicolumn{1}{c}{\textbf{}} \\ \hline \\
\endfirsthead

\multicolumn{8}{c}%
{{Table \thetable{} -- continued from previous page}} \\
\hline \\
\multicolumn{1}{c}{\textbf{Star Id.}} &
\multicolumn{1}{c}{\textbf{$T_{eff}$}} & 
\multicolumn{1}{c}{\textbf{$\log(g)$}} &
\multicolumn{1}{c}{\textbf{$v\,\sin(i)$}} &
\multicolumn{1}{c}{\textbf{$\xi$}} &
\multicolumn{1}{c}{\textbf{$\log(\frac{C}{H})+12$}} &
\multicolumn{1}{c}{\textbf{$\log(\frac{N}{H})+12$}} &
\multicolumn{1}{c}{\textbf{ref}} \\
\multicolumn{1}{c}{\textbf{}} &
\multicolumn{1}{c}{\textbf{(K)}} & 
\multicolumn{1}{c}{\textbf{}} &
\multicolumn{1}{c}{\textbf{(km s$^{-1}$)}} &
\multicolumn{1}{c}{\textbf{(km s$^{-1}$)}} &
\multicolumn{1}{c}{\textbf{}} &
\multicolumn{1}{c}{\textbf{}} &
\multicolumn{1}{c}{\textbf{}} \\ \hline \\
\endhead

\hline \multicolumn{8}{r}{{Continued on next page}} \\ \hline
\endfoot

\hline \\ \hline \multicolumn{8}{l}{{References: TW = This Work; AND99 = \citet{Andrievsky1999}.}}\\
\endlastfoot
HD32249 & 20000 $\pm$ 130 & 4.26 $\pm$ 0.13 & \phantom{11}86 $^{+11}_{-7}$ & \phantom{1111}5 $\pm$ 1 & 8.19 & 7.8 & TW \\*
& \textit{18500 $\pm$ 1000} & \textit{4. $\pm$ 0.3} & \phantom{11}\textit{48 $\pm$ 8} & \phantom{1111}\textit{4 } &  &  & \textit{AND99} \\[4pt]
HD35912 & 19610 $^{+140}_{-240}$ & 4.27 $\pm$ 0.04 & \phantom{11}17 $^{+15}_{-17}$ & \phantom{1111}1 $^{+2}_{-1}$ & 8.28 & 7.76 & TW \\*
& \textit{19000 $\pm$ 1000} & \textit{4.1 $\pm$ 0.3} & \phantom{11}\textit{22 $\pm$ 4} & \phantom{1111}\textit{4 } &  &  & \textit{AND99} \\[4pt]
HD886 & 21250 $\pm$ 130 & 3.97 $\pm$ 0.04 & \phantom{111}4 $^{+16}_{-4}$ & \phantom{1111}8 $\pm$ 1 & 8.11 & 7.55 & TW \\*
& \textit{21600 $\pm$ 1000} & \textit{3.7 $\pm$ 0.3} & \phantom{111}\textit{8 $\pm$ 2} & \phantom{1111}\textit{3.3 } &  &  & \textit{AND99} \\
\end{longtable}

\end{center}

\begin{center}
\begin {longtable}{clll@{}l@{}c}
\caption[Basic parameter field]{Basic Parameters of Field B Stars Returned by ATLAS9 with Solar Composition from \citet{Grevesse1998}} \label{tab:pftabBK} \\

\hline \multicolumn{1}{c}{\textbf{Star Id.}} &
\multicolumn{1}{c}{\textbf{$T_{eff}$}} & 
\multicolumn{1}{c}{\textbf{$\log(g)$}} &
\multicolumn{1}{c}{\textbf{$v\,\sin(i)$}} &
\multicolumn{1}{c}{\textbf{$\xi$}} &
\multicolumn{1}{c}{\textbf{ref}} \\
\multicolumn{1}{c}{\textbf{}} &
\multicolumn{1}{c}{\textbf{(K)}} & 
\multicolumn{1}{c}{\textbf{}} &
\multicolumn{1}{c}{\textbf{(km s$^{-1}$)}} &
\multicolumn{1}{c}{\textbf{(km s$^{-1}$)}} &
\multicolumn{1}{c}{\textbf{}} \\ \hline \\
\endfirsthead

\multicolumn{6}{c}%
{{Table \thetable{} -- continued from previous page}} \\
\hline \multicolumn{1}{c}{\textbf{Star Id.}} &
\multicolumn{1}{c}{\textbf{$T_{eff}$}} & 
\multicolumn{1}{c}{\textbf{$\log(g)$}} &
\multicolumn{1}{c}{\textbf{$v\,\sin(i)$}} &
\multicolumn{1}{c}{\textbf{$\xi$}} &
\multicolumn{1}{c}{\textbf{ref}} \\
\multicolumn{1}{c}{\textbf{}} &
\multicolumn{1}{c}{\textbf{(K)}} & 
\multicolumn{1}{c}{\textbf{}} &
\multicolumn{1}{c}{\textbf{(km s$^{-1}$)}} &
\multicolumn{1}{c}{\textbf{(km s$^{-1}$)}} &
\multicolumn{1}{c}{\textbf{}} \\ \hline \\
\endhead

\hline \multicolumn{6}{r}{{Continued on next page}} \\ \hline
\endfoot

\hline \\ \hline \multicolumn{6}{l}{{References: TW = This Work; TWHG = This Work using H$\gamma$ only; }}\\
\multicolumn{6}{l}{{HG10 = \citet{Huang2010}. }}\\
\multicolumn{6}{l}{{Numbers between parenthesis refer to the 'Star id n\textdegree' in Figure \ref{fig:Delta_teff_logg}. }} \\ 
\endlastfoot
HD214263 & 19760 $^{+120}_{-140}$ & 4.08 $\pm$ 0.04 & \phantom{11}88 $^{+7}_{-8}$ & \phantom{1111}1 $^{+2}_{-1}$ & TW \\*
 (16) & \textit{18570 $\pm$ 350} & \textit{3.87 $\pm$ 0.06} & \phantom{11}\textit{63} & \phantom{1111}\textit{2 } & \textit{TWHG} \\*
& \textit{19027 $\pm$ 250} & \textit{3.807 $\pm$ 0.025} & \phantom{11}\textit{63 $\pm$ 16} & \phantom{1111}\textit{2 } & \textit{HG10} \\[4pt]
HD214432 & 18870 $^{+130}_{-120}$ & 4.47 $\pm$ 0.04 & \phantom{1}132 $^{+8}_{-7}$ & \phantom{1111}5 $\pm$ 1 & TW \\*
 (17) & \textit{15560 $^{+450}_{-560}$} & \textit{3.95 $\pm$ 0.08} & \phantom{1}\textit{137} & \phantom{1111}\textit{2 } & \textit{TWHG} \\*
& \textit{16124 $\pm$ 200} & \textit{3.890 $\pm$ 0.028} & \phantom{1}\textit{137 $\pm$ 9} & \phantom{1111}\textit{2 } & \textit{HG10} \\[4pt]
HD215191 & 20500 $^{+130}_{-120}$ & 3.91 $\pm$ 0.04 & \phantom{1}200 $\pm$ 5 & \phantom{1111}6 $\pm$ 1 & TW \\*
 (18) & \textit{20490 $\pm$ 360} & \textit{3.88 $\pm$ 0.06} & \phantom{1}\textit{195} & \phantom{1111}\textit{2 } & \textit{TWHG} \\
\end{longtable}

\end{center}

\begin{center}
\begin{longtable}{cclll@{}l@{}c}
\caption[stellar parameters cluster BK]{Stellar Parameters of Cluster B Stars Returned by ATLAS9 with Solar Composition from \citet{Grevesse1998}} \label{tab:pftabclBK} \\

\hline \multicolumn{1}{c}{\textbf{Cluster Id.}} &
\multicolumn{1}{c}{\textbf{Star Id.}} &
\multicolumn{1}{c}{\textbf{$T_{eff}$}} & 
\multicolumn{1}{c}{\textbf{$\log(g)$}} &
\multicolumn{1}{c}{\textbf{$v\,\sin(i)$}} &
\multicolumn{1}{c}{\textbf{$\xi$}} &
\multicolumn{1}{c}{\textbf{ref}} \\
\multicolumn{1}{c}{\textbf{}} &
\multicolumn{1}{c}{\textbf{}} &
\multicolumn{1}{c}{\textbf{(K)}} & 
\multicolumn{1}{c}{\textbf{}} &
\multicolumn{1}{c}{\textbf{(km s$^{-1}$)}} &
\multicolumn{1}{c}{\textbf{(km s$^{-1}$)}} &
\multicolumn{1}{c}{\textbf{}} \\ \hline \\
\endfirsthead

\multicolumn{7}{c}%
{{Table \thetable{} -- continued from previous page}} \\
\hline \multicolumn{1}{c}{\textbf{Cluster Id.}} &
\multicolumn{1}{c}{\textbf{Star Id.}} &
\multicolumn{1}{c}{\textbf{$T_{eff}$}} & 
\multicolumn{1}{c}{\textbf{$\log(g)$}} &
\multicolumn{1}{c}{\textbf{$v\,\sin(i)$}} &
\multicolumn{1}{c}{\textbf{$\xi$}} &
\multicolumn{1}{c}{\textbf{ref}} \\
\multicolumn{1}{c}{\textbf{}} &
\multicolumn{1}{c}{\textbf{}} &
\multicolumn{1}{c}{\textbf{(K)}} & 
\multicolumn{1}{c}{\textbf{}} &
\multicolumn{1}{c}{\textbf{(km s$^{-1}$)}} &
\multicolumn{1}{c}{\textbf{(km s$^{-1}$)}} &
\multicolumn{1}{c}{\textbf{}} \\ \hline \\
\endhead

\hline \multicolumn{7}{r}{{Continued on next page}} \\ \hline
\endfoot

\hline \\ \hline
\multicolumn{7}{l}{{References: TW = This Work; TWHG = This Work using H$\gamma$ only;}} \\ 
\multicolumn{7}{l}{{HG10 = \citet{Huang2010}; and HG06 = \citet{Huang2006a,Huang2006b}. }} \\ 
\multicolumn{7}{l}{{Numbers between parenthesis refer to the 'Star id n\textdegree' in Figure \ref{fig:Delta_teff_logg}. }} \\ 
\endlastfoot

NGC1960  & BD341103s & 18870 $^{+130}_{-120}$ & 4.31 $\pm$ 0.04 & \phantom{1}290 $^{+9}_{-6}$ & \phantom{1111}2 $\pm$ 1 & TW \\*
\textit{} & (0) & \textit{20030 $\pm$ 920} & \textit{4.40 $\pm$ 0.17} & \phantom{11}\textit{77} & \phantom{1111}\textit{2 } & \textit{TWHG} \\*
\textit{} & \textit{} & \textit{17297 $\pm$ 150} & \textit{3.983 $\pm$ 0.015} & \phantom{11}\textit{77 $\pm$ 7} & \phantom{1111}\textit{2 } & \textit{HG10} \\[4pt]
& SAB20 & 12970 $^{+150}_{-250}$ & 3.87 $^{+0.05}_{-0.07}$ & \phantom{1}274 $^{+14}_{-10}$ & \phantom{1111}1 $^{+2}_{-1}$ & TW \\*
\textit{} & (8) & \textit{14560 $\pm$ 620} & \textit{4.25 $\pm$ 0.15} & \phantom{1}\textit{141} & \phantom{1111}\textit{2 } & \textit{TWHG} \\*
\textit{} & \textit{} & \textit{14845 $\pm$ 100} & \textit{4.242 $\pm$ 0.018} & \phantom{1}\textit{141 $\pm$ 7} & \phantom{1111}\textit{2 } & \textit{HG10} \\[4pt]
& SAB22 & 14500 $\pm$ 130 & 4.33 $\pm$ 0.04 & \phantom{1}126 $^{+10}_{-5}$ & \phantom{1111}3 $\pm$ 1 & TW \\*
\textit{} & (9) & \textit{13500 $\pm$ 210} & \textit{4.09 $\pm$ 0.06} & \phantom{1}\textit{135} & \phantom{1111}\textit{2 } & \textit{TWHG} \\
\end{longtable}

\end{center}

\label{lastpage}

\end{document}